\documentclass[12pt,preprint]{aastex}
\usepackage{natbib}
\usepackage{bs}

\def\ms{m s$^{-1}$}
\def\hatalpha{\boldsymbol{\hat{\alpha}}}

\shorttitle{The HARPS-Terra project}
\shortauthors{Anglada-Escud\'e \& Butler}

\begin{document}

\title{The HARPS-TERRA project I.
Description of the algorithms, performance
and new measurements on a few remarkable
stars observed by HARPS.}

\author{
Guillem Anglada-Escud\'e\altaffilmark{1,2}
R. Paul Butler\altaffilmark{1}
}

\email{
anglada@dtm.ciw.edu, }

\altaffiltext{1}{Carnegie Institution of Washington, Department of Terrestrial
Magnetism, 5241 Broad Branch Rd. NW, Washington D.C., 20015, USA}
\altaffiltext{2}{Universit\"{a}t G\"{o}ttingen, Institut f\"ur Astrophysik, Friedrich-Hund-Platz 1,37077 G\"{o}ttingen, Germany}

\begin{abstract} Doppler spectroscopy has uncovered or
confirmed all the known planets orbiting nearby stars.
Two main techniques are used to obtain precision Doppler
measurements at optical wavelengths. The first approach
is the gas cell method, which consists on the
least-squares matching of the spectrum of Iodine
imprinted on the spectrum of the star. The second method
relies on the construction of a stabilized spectrograph
externally calibrated in wavelength. The most precise
stabilized spectrometer in operation is HARPS, operated
by ESO in La Silla Observatory/Chile. The Doppler
measurements obtained with HARPS are typically obtained
using the Cross-Correlation Function  technique (CCF).
It consists of multiplying the stellar spectrum with a
weighted binary mask and finding the minimum of such
product as a function of the Doppler shift. It is known
that CCF is suboptimal in exploiting the Doppler
information in the stellar spectrum. Here, we describe
an algorithm to obtain precision RV measurements using
least-squares matching of each observed spectrum to a
high signal-to-noise ratio template derived from the
same observations. Such algorithm is implemented in our
software called HARPS-TERRA (Template Enhanced Radial
velocity Re-analysis Application). New radial velocity
measurements on a representative sample of stars
observed by HARPS is used to illustrate the benefits of
the proposed method. We show that, compared to CCF,
template matching provides a significant improvement in
accuracy, specially when applied to M dwarfs. 
\end{abstract}


\section{Introduction}

The Doppler technique has been the most
successful method to detect and confirm the
presence of extrasolar planets around nearby
stars \citep{mayor:1995,marcy:1996}. The
stability of the spectrographs and the data
analysis techniques used to obtain precision
radial velocity (RV) measurements have been
steadily improving during the last 16 years
of exoplanet discoveries. In particular, the
vast majority of candidates detected via
Doppler spectroscopy have been obtained
using two approaches : the gas cell method
and the stabilized spectrograph approach.

Because an absorption cell can be installed at
low cost on any general purpose echelle spectrometer,
it is the most broadly used technique. In this
approach, a glass cell filled with Iodine gas at
low pressure is inserted just before the entrance
slit of the spectrometer. Because the stellar
spectrum is imprinted with Iodine prior to
entering the spectrograph, the cell spectrum
tracks the same instrumental distortions suffered
by the stellar lines. Thanks to this, the
wavelength solution and the instrumental profile
can be fitted simultaneously to the Doppler shift
of the star allowing nominal RV precisions at the
level of 1-2 \ms. The data analysis method
required to extract such precise measurements
consists in forward-modeling the spectrum of
Iodine multiplied to a high SNR template of the
star, and convolving the product with a
parameterized model of the instrumental profile.
This method was pioneered by \citet{butler:1996}
and has been used to infer the presence of more
than 300+ exoplanets. Due to the large number of
parameters involved, this technique is
computationally very intensive. Also, it requires
a high SNR realization of the real stellar
spectrum at higher resolution than the actual
observations. Because obtaining a high
resolution perfectly calibrated spectrum of the
star is usually not possible, obtaining a
realistic template from the deconvolution of
observations without Iodine is a key
element of the method and is thought to be one
of its major limitations \citep[e.g., see
supplementary material in][]{howard:2010}.

The other leading method to obtain precision radial
velocity measurements consists of building a fiber
fed, very stable spectrograph which is then
wavelength calibrated using an external source
such as a Th/Ar emission lamp. This technique has
been developed and refined by the group lead by M.
Mayor (from now on the Geneva group) over the past
20 years. This approach provided the first clear
detection of an extrasolar planet around a
solar-like star \citep{mayor:1995} using the
ELODIE spectrograph at Observatoire de
Haute-Provence/France. The interested reader can
find the basic elements in the construction of a
fiber-fed stabilized spectrograph in
\citet{baranne:1996}. While the gas cell technique
requires the simultaneous adjustment of the
wavelength scale, instrumental profile and Doppler
offset of the star; the stabilized spectrograph
approach allows tackling each problem separately.
First, the instrumental profile is constant by
design thanks to the use of an image scrambling
system coupled with the optical fibers. In the
design by \citet{baranne:1996}, the wavelength
calibration is obtained using two optical fibers
closely packed together at the entrance of the
spectrometer. Therefore, they follow almost
identical optical paths within the spectrometer
optics tracking almost the same optical
distortions. The first fiber (science fiber) is
illuminated with a wavelength calibration source
(e.g., Th/Ar lamp) at the beginning of the night
and the nominal absolute wavelength solution
relative to that source is obtained
\citep{pepe:2002}. The second fiber is fed with
the same calibration source during the calibration
of the science fiber and during the science observations.
Although there might be a significant RV offset
between fibers, they are similarly affected by
changes in the instrument and, as a consequence,
they share similar intra-night wavelength drifts.
This second fiber is only used to monitor
intra-night changes in the instrument and the
information it provides is an RV drift to be added
to the measured RVs.
Finally, the RV measurement on a fully calibrated
spectrum is obtained by the so-called
Cross-Correlation Function method (or CCF). CCF is
based on multiplying the observed spectrum by a
weighted binary mask. The binary mask is different
from zero on the nominal positions of the stellar
lines, and each non-zero chunk is weighted
according to the relative depth of the stellar
line against the local continuum. This binary mask
is then Doppler shifted and the CCF is evaluated
again. The minimum of the CCF as a function of the
Doppler offset is the desired RV measurement. To
achieve higher precision, the shape of the CCF is
centroided using a Gaussian profile.
The CCF method and some details on how the binary masks
are obtained is outlined in
\citet{queloz:1995} and \citet{pepe:2002}.

The flagship stabilized spectrometer build
by the Geneva group is HARPS (High Accuracy
Radial velocity Planet Searcher) installed
in the 3.6 m Telescope at the European
Southern Observatory (ESO) site in La
Silla/Chile \citep{harps:construction}.
Thanks to a careful design and construction
(vacuum sealed tank, high mechanical
stability and accurate temperature
control), the wavelength solution and
instrumental profile are very stable. For
example, intra-night Doppler drifts as
measured by the calibration fiber are
typically smaller than 0.5 \ms. Long
exposures (t$>$200 sec) using the
calibration fiber cause leaks of the Th/Ar
spectrum on the science spectrum, so the
calibration fiber is only used when
observing very bright targets and when extreme RV
precision is required. The list of planets
detected by HARPS is long an varied as can
be seen in the 34 papers of the series
\textit{The HARPS search for southern
extra-solar planets}. Instead of citing all
of them, we refer the interested reader to
the latest HARPS results presented in
\citet{pepe:2011,mayor:2011}. HARPS has
demonstrated an RV stability at the level of
1 \ms\ on time-scales of several years.

Despite these impressive results, it is known that
the CCF method implemented in the HARPS Data
Reduction Software is suboptimal in the sense that
it does not exploit the full Doppler information
on the stellar spectrum \citep[e.g. see
][]{queloz:1995, pepe:2002}. We asked ourselves if
a least squares approach where the observed
stellar spectrum is matched to a high SNR template could be
used to extract higher RV precision on HARPS
observations. Given that a number of stabilized
spectrographs are under construction (e.g.
HARPS-North, ESPRESSO/ESO, Carmenes/CaHa), and
that there is significant investment in hardware
development to achieve higher RV stability, it is
important that the used data analysis methods are
as optimal as possible. Also, precision RVs are
difficult to reproduce and, given that the stakes
are ambitious (detection of potentially habitable
worlds), the use of different RV measurement
improves the detection confidence of low amplitude signals. In
this work we derive from first principles the
algorithms of the template matching
technique to be applied on stabilized
spectrographs and implement it to public HARPS
observations. HARPS is an ESO instrument and, as
such, all the data obtained from it becomes
publicly available after a proprietary period of a
few months (or years). Since January 2011, the
data products derived from the HARPS-ESO Data
Reduction Software (wavelength calibrated spectra,
but also CCF measured RVs among others) are
publicly available through a dedicated webpage in
the ESO website
\footnote{\texttt{http://archive.eso.org/wdb/wdb/eso/repro/form}}.
All the HARPS data used in this work has been
obtained from there.

The core of our project is our software tool called
HARPS-TERRA, where TERRA stands for
\textit{Template Enhanced Radial velocity
Re-analysis Application}. HARPS-TERRA handles the
full process of unpacking the HARPS-ESO archive
files, generation of a high SNR template and the
obtention of the final RV measurement in a single
command line call. This software is custom made,
fully coded in Java, and can run on any machine
supporting a Java Run-time Environment 1.6 or
newer.

In Section \ref{sec:algorithms}, we derive and
describe the basic algorithms used to obtain RV
measurements from HARPS reduced spectra. In
Section \ref{sec:parameters}, we investigate the
optimal RV extraction parameters and define the
\textit{standard setup} to obtain optimal RV
precision with minimum human intervention on G, K
and M dwarfs. Section \ref{sec:comparison} shows
the performance of HARPS-TERRA on a few
representative data sets; demonstrating a
significant improvement in precision, especially
on M dwarfs. In the same section, we use new RV
measurements obtained with HARPS-TERRA to discuss
planetary systems proposed on a few
representative stars observed by HARPS.

HARPS-TERRA is still in development but can
be distributed for particular applications
upon request. Given that other HARPS
programs could benefit from using it (e.g.
asteroseismology programs, binary stars), we
plan a public release of the tool in the
near future.


\section{Description of the algorithms} \label{sec:algorithms}

The proposed algorithm is based on
minimizing the differences of the observed
spectrum against a parameterized template. In our
particular implementation, we first use the higher
SNR observation as a preliminary template. In a
second iteration, a very high SNR template is
obtained by coadding all the observations and the
RVs are measured again. Note that the template
will be already convolved with the instrumental
profile so this method (as the CCF technique does)
relies on the long term stability of the
instrumental profile and wavelength calibration
strategy.

HARPS is a cross-dispersed echelle spectrograph.
Therefore, the stellar spectrum is split in
diffraction orders over the detector (also called
\textit{echelle apertures} , or
\textit{apertures}). Each aperture has to be
extracted very carefully in a complex process from
the raw CCD images. The wavelength calibration is
usually made on a nightly basis using a standard
set of calibration frames (i.e., flat, darks, Th/Ar
lamps) that are taken once at the
beginning of the night. Thankfully, all this
extraction and the corresponding wavelength
solution is efficiently implemented by the
HARPS-ESO Data Reduction Software (hereafter HARPS
DRS) developed by the HARPS-ESO
team\footnote{http://www.eso.org/sci/facilities/lasilla/instruments/harps/doc/index.html},
and will not be discussed here. The wavelength
calibrated spectra are provided in conveniently
formatted 'fits' files described in the HARPS DRS
manual. As of November 2011, HARPS-TERRA is
designed to work with the output files of the
HARPS DRS v3.5, but should be easily adapted to
future updates of the HARPS DRS and/or other stabilized
spectrographs. An overview of the HARPS spectral
format is given in Table \ref{tab:harpsparams}. As
described in the introduction, the secondary
calibration fiber is typically only used on bright
stars to achieve maximum precision. If the
observer used this secondary fiber, such drift is
also provided by the HARPS DRS and will be added
to the final RV measurement of each echelle
aperture. In all that follows we treat each
echelle aperture as an independent spectrum and
the final RV measurement will be a weighted mean
of the RVs measured across all such apertures.

\begin{deluxetable}{lr}  
\tabletypesize{\small}
\tablecaption{HARPS spectral parameters relevant to this work}
\tablehead{
  \colhead{Parameter} &
  \colhead{Value} \\
}
\startdata
Spectral resolution $\lambda/\delta \lambda$ at 5500\AA & 120 000 \\
Number of diffraction orders & 72 \\
(or echelle apertures)  \\
Pixels in each aperture & 4096\\
Wavelength range (Blue CCD) & 3780 -- 5300 \AA \\
Wavelength range (Red CCD) & 5330 --6910 \AA \\
Sampling per\\
resolution element & 4.1 pix\\
\enddata
\label{tab:harpsparams}
\end{deluxetable}

Let us assume that we have a very high SNR,
wavelength calibrated spectrum of the star (template).
Given a wavelength calibrated observation and
a stable instrumental profile, the observed
spectrum differs from the template only by a
Doppler shift (due to Earth motion around the
Sun and/or the presence of companions) and a
flux normalization function across each
aperture. We observed that this flux
normalization is time dependent at the few \%
level most likely due to observational and
instrumental effects, such as atmospheric
differential refraction, differential
absorption or telescope tracking errors. Let
us define the difference $R$ between the
template $F$ and the observed flux $f$ at each
wavelength $\lambda$ as
\begin{eqnarray}
R\left[\lambda;\hatalpha\right]
  = F\left[ \alpha_v \lambda\right]-
         f\left[\lambda\right]
	 \sum_{m=0}^{M}
         \alpha_m
	 \left(\lambda-\lambda_c\right)^{m}
\label{eq:residual}
\end{eqnarray}

\noindent where $\lambda$ is the wavelength of the
observation transformed to the Solar System
barycenter reference frame. The set of free
parameters is represented by $\hatalpha =
\left[\alpha_v, \alpha_0, \ldots,
\alpha_M\right]$. The first term on the right side
is the template evaluated at $\alpha_v \lambda_i$,
where $\alpha_v$ is the Doppler factor on which we
are interested. To derive a differential RV
measurement from this Doppler factor, we use the
very simple expression
$\alpha_v=1-v_r/c$ \citep{stumpff:1979}, where $c$
is the speed of light and $v_r$ is the relative
radial velocity between the observer and the star
on which we are interested. The second term is the
observed flux $f$ multiplied by an M-degree polynomial accounting for the flux normalization across each
aperture (also called blaze function correction),
and $\lambda_c$ is the central wavelength of a
given aperture. Note that we could apply this
flux normalization to the template $F$ instead to
the observed flux $f$. However, this would couple the
flux normalization coefficients $\alpha_m$ to the
Doppler factor $\alpha_v$ in a non-linear fashion
which, from a numerical point of view, is an
undesirable complication.

Now this difference $R$ can be Taylor expanded
around some nominal values for the parameters
$\hatalpha_{(0)}$ in powers of the parameter
increments $\delta\hatalpha$ as
\begin{eqnarray}
R\left[\lambda;\hatalpha\right] &\simeq&
R\left[\lambda;\hatalpha_{(0)}\right] +
\left.
    \frac{\partial R}{\partial \alpha_v}
\right|_{\lambda;\,\hatalpha_{(0)}}
\delta\alpha_v +
\sum_{m=0}^{M}
\left.
\frac{\partial R}{\partial \alpha_m}
\right|_{\lambda;\,\hatalpha_{(0)}}
\delta\alpha_m
\\
\left.
\frac{\partial R}{\partial \alpha_v} 
\right|_{\lambda;\,\hatalpha_{(0)}}
&=&
\left.
\lambda \frac{d F}{d \left(\alpha_v\lambda\right)}
\right|_{\alpha_{v(0)}\lambda}
\label{eq:partialv}
\\
\left.
\frac{\partial R}{\partial \alpha_m}
\right|_{\lambda;\,\hatalpha_{(0)}}
\,\, &=& \,\,
-f\left[\lambda\right]\left(\lambda-\lambda_c\right)^m \,.
\label{eq:partiala}
\end{eqnarray}
\noindent The weighted sum of these R over
all the observed wavelengths $\lambda_i$ (or
pixels) is defined as
\begin{eqnarray}
\chi^2 = \sum^{N_{pix}}_{i=1}
\omega_i R\left[\lambda_i;\hatalpha
\right]^2
\, .\label{eq:chi2}
\end{eqnarray}
\noindent and is the quantity to be minimized. The partial derivative of $\chi^2$ with
respect to each increment $\delta \alpha_v,
\delta \alpha_0,\delta \alpha_1,\ldots\delta
\alpha_M,$ equated to 0 generates the a system of equations for
such increments (the so-called normal equations)
which can then can be solved using standard matrix
techniques. The quantities $\omega_i$ are the
weights assigned to each $\lambda_i$ and their
precise value will be discussed later. The
resulting system of equations reads
\begin{eqnarray}
\sum_{i}^{N_{pix}}
\left.
\omega_i
\frac{\partial R}{\partial \alpha_v}
\frac{\partial R}{\partial \alpha_v}
\right|_{\lambda_i;\,\hatalpha_{(0)}}
\, \delta \alpha_v +
\ldots +
\sum_{i}^{N_{pix}}
\left.
\omega_i
\frac{\partial R}{\partial \alpha_v}
\frac{\partial R}{\partial \alpha_M}
\right|_{\lambda_i;\,\hatalpha_{(0)}}
\, \delta \alpha_M
&=&
-\sum_{i}^{N_{pix}}
\left.
\omega_i
R
\frac{\partial R}{\partial \alpha_v}
\right|_{\lambda_i;\,\hatalpha_{(0)}}
\label{eq:normal0}
\\
%
%
\ldots\nonumber \\
\ldots\nonumber \\
%
%
\sum_{i}^{N_{pix}}
\left.
\omega_i
\frac{\partial R}{\partial \alpha_M}
\frac{\partial R}{\partial \alpha_v}
\right|_{\lambda_i;\,\hatalpha_{(0)}}
\, \delta \alpha_v +
\ldots +
\sum_{i}^{N_{pix}}
\left.
\omega_i
\frac{\partial R}{\partial \alpha_M}
\frac{\partial R}{\partial \alpha_M}
\right|_{\lambda_i;\,\hatalpha_{(0)}}
\, \delta \alpha_M
&=&
-\sum_{i}^{N_{pix}}
\left.
\omega_i
R
\frac{\partial R}{\partial \alpha_M}
\right|_{\lambda_i;\,\hatalpha_{(0)}}
\label{eq:normalM}
%
%
\end{eqnarray}
\noindent where the values of $R$ are obtained
using Equation \ref{eq:residual} and the
partial derivatives are computed using Equations
\ref{eq:partialv} and \ref{eq:partiala}.
Because the Doppler factor $\alpha_v$ is
nonlinear, this system of equations has to be
solved iteratively. Each iteration consists in
1) compute the normal equations using the
current values of the parameters
$\hatalpha_{(0)}$, 2) solve for the
parameter increments $\delta\hatalpha$, 3)
update the parameter values
$\hatalpha_{new} =
\hatalpha_{0}+\delta\hatalpha$. These
equations can be written in matrix form as
\begin{eqnarray}
A_{lk}\delta\alpha_{k} = b_l\, ,
\end{eqnarray}
\noindent where $A_{lk}$ is the M+1 $\times$
M+1 matrix of coefficients multiplying the
parameter increments $\delta\alpha_{k}$ in the normal
equations \ref{eq:normal0}--\ref{eq:normalM}. $A_{lk}$ 
is equivalent to the Hessian matrix of the
$\chi^2$ and sometimes is called curvature
matrix. By straight-forward error propagation
\citep{numerical} one finds that, when the
solution converges to the $\chi^2$ minimum,
the formal uncertainties in each free parameter can
be derived from the inverse of the $\mathbf{A}$ matrix
as

\begin{eqnarray}
  \sigma_{\alpha_k} = \sqrt{\left(\mathbf{A}^{-1}\right)_{kk}}\, .
  \label{eq:formalerrors}
\end{eqnarray}

\noindent where $\mathbf{A}^{-1}$ is usually called 
the \textit{covariance matrix}. Even though this prescription ignores
correlation between parameters, it still provides
a good estimate of the formal precision of the
measurement if the weights $\omega_i$ are properly
estimated. The formal uncertainty in the Doppler coefficient
$\delta\alpha_v$ multiplied by the speed of light
($\sigma_v$ = c $\delta\alpha_v $) is used to compute the initial 
weight of each echelle aperture in the computation of the mean epoch RV.
Therefore one needs to be sure that these formal
uncertainties are as realistic as possible (see
sub-section \ref{sec:weights})

The radial velocity measurement obtained on each
order, its formal uncertainty and other
information about the fit (flux normalization
coefficients, number of iterations, number of
masked pixels, RMS of the fit, etc.) are stored in
a file for further processing (see Section
\ref{sec:finalrv}). At this point, no assumption
is done on the nature of the star and all the
apertures are processed irrespective of their SNR
or overall quality.

\begin{figure*}[tb]
   \centering
   \includegraphics[width=5.0in,clip]{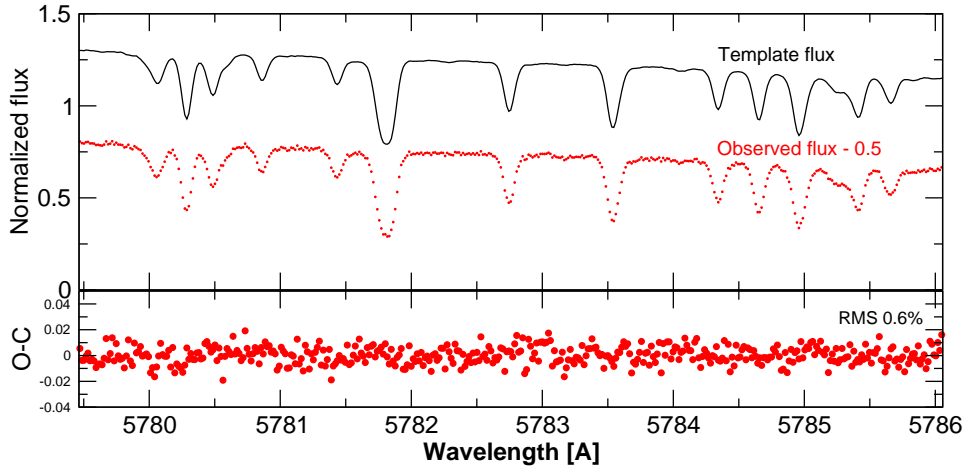}

\caption{Result of the fit to a small chunk of
the spectrum of Tau Ceti (G8.5V). The average
SNR of the observation is 180. Such SNR would
correspond to an RMS of 0.55\% if only photon
noise were involved. The actual RMS is 0.65\%.
Even though it is a very good fit, this slight
excess of RMS illustrates that assuming
Poisson statistics is usually too optimistic
to obtain the weight of each pixel and derive
a realistic formal RV precision. A cubic flux
normalization polynomial has been applied to
match the blaze function of the observations
to the blaze function of the template. The
nominal RV uncertainty for this aperture
(aperture 54) is 5.4 \ms.}

\label{fig:spectrumfitG8}
\end{figure*}

\begin{figure*}[tb]
   \centering
   \includegraphics[width=5.0in,clip]{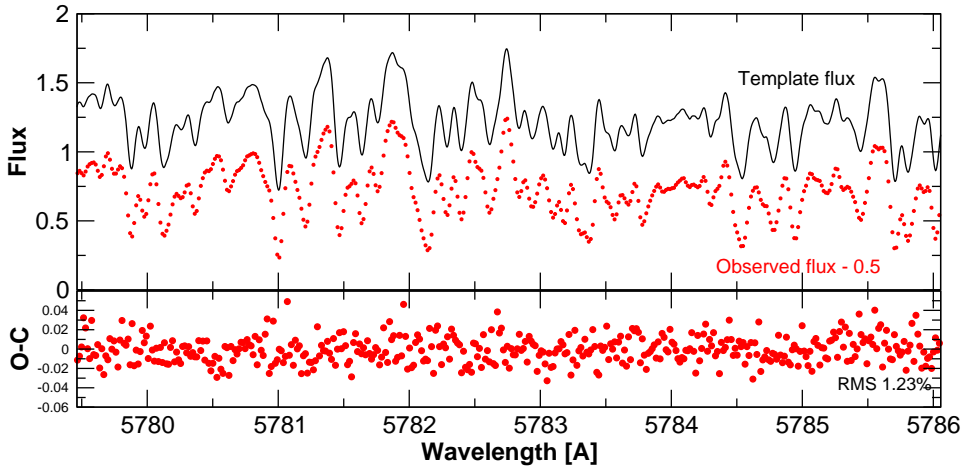}

\caption{Results of the template matching
process to a small chunk of the spectrum of GJ
699 (M4V). The average SNR of the observation
is 83 which should correspond to an RMS of
1.22\% compared to the actual 1.23\% obtained.
Even though the SNR is almost a third compared
to the SNR of the G dwarf in Figure
\ref{fig:spectrumfitG8}, the much more
abundant Doppler information in the spectral
features of the M dwarf gives a nominal
uncertainty of 2.68 \ms for this observation.}

\label{fig:spectrumfitM4}
\end{figure*}

\subsection{Algorithm implementation details}

Even though the basic algorithm has already be
outlined, a number of technical details must
be carefully addressed to reach maximal RV
precision in an efficient numerical fashion.
Here we describe those technical aspects that
require special attention and provide the
practical solutions we have adopted.

\subsubsection{Template interpolation}

Until now, we have assumed a theoretical
continuous and differentiable function for the
template. In reality, we need to build the
template from the same observations. Since the
observed wavelengths do not necessarily coincide
with the wavelengths where the template is
sampled, $F$ and its derivatives must be obtained
through interpolation. In our implementation, the
interpolation of the discretely sampled template
is obtained using a cubic spline
\citep{numerical}. Cubic splines have the property
of producing a continuous version of the
derivative, which is critical for the convergence
properties of the non-linear least-squares
algorithm. The details on how such a template is
generated are given in Section \ref{sec:template}.

\subsubsection{Pixel weighting}\label{sec:weights}

Each weight $\omega_i$ should be a function of the
uncertainty in $R_i$. Assuming a very high SNR
template and Poisson statistics, the uncertainty
in the flux of each pixel is just
$\sigma_i=\sqrt{f_i}$. However, we found that
assuming such uncertainty produced suboptimal
results because (1) it tends to overweight pixels
with high flux and little real Doppler information
(continuum), and (2) Poison statistics never
applies in realistic situations (e.g., pixel
non-linearities, imperfect flat fielding, unmasked
telluric features, imperfect template). Instead,
we re-scale the Poisson uncertainty of each pixel
using an empirical approach. In the first
iteration of equations
\ref{eq:normal0}--\ref{eq:normalM}, Poisson
statistics are assumed on each pixel so
$\omega_{i(0)}=\sqrt{f_i}$. After that, we
compute the RMS of the $R_i$ differences over all
the pixels and we call it
$\sigma_{\left<R\right>}$. If Poisson statistics
applied, $\sigma_{\left<R\right>}$ should be equal
to the square root of the mean pixel flux
($\sqrt{\left<f\right>}$). However, if there are
additional sources of noise
$\sigma_{\left<R\right>}/\sqrt{\left<f\right>}=
\kappa$, where $\kappa>1$. The weights to be used
in the next iteration are therefore re-scaled with
this quality factor $\kappa$ as $\omega_{i(new)} =
\left(\kappa^2 f_i\right)^{-1}$. The $\kappa$
value of the last least-squares iteration is also
stored in a file and provides useful information
to assess the level of systematic noise on each
echelle order.

\subsubsection{Telluric masking and outlier filtering}

The weights are also used to mask those pixels
with undesirable properties. This is, any
$\omega_i$ on wavelengths coincident with
telluric absorption features deeper than 1\% is
set to 0. A synthetic spectrum of the atmosphere
was used to identify such telluric lines and
generate a list of wavelengths (or telluric mask)
to be avoided. The same telluric mask is applied
to all the observations.

Because of the Barycentric motion of the Earth
($\sim \pm 15$ k\ms), the stellar spectrum at the
border of each echelle order is not always present
in all the epochs. This going in-and-out of the
borders can generate small
systematic RV offsets correlated with the Earth
barycentric motion. Even though a more
sophisticated solution could be adopted and for
the sake of simplicity, the weights of pixels
within 0.8 \AA\ of each aperture border are also
set to 0. Although some Doppler information is
lost in the process, the flux at the borders is
significantly lower than at the center of the
apertures and some extra random noise is
preferred over correlated noise with a 1 year
periodicity. We tested less restrictive border
masking obtaining almost identical results.

After the first least squares iteration is
obtained, the weights of possible outliers are
also set to 0. A pixel is considered an
outlier if it has a residual 4 times larger
than the empirically determined
$\sigma_{\left<R\right>}$ (i.e., 4-$\sigma$
clipping). This threshold is arbitrary but we
found it does a good job removing pixel
outliers (e.g., cosmic ray hits), while
preserving most of the well behaved ones.

\subsubsection{Parameter initialization and
barycentric correction}

The described algorithm is a non-linear least
squares solver and, as such, it requires an
initial guess for the values of the free
parameters. The initial value for the flux
polynomial coefficient $\alpha_0$ is set to
the mean pixel flux $\left<f\right>$ and all
the other $\alpha_m$ are set to 0. Note that
the real flux normalization function (or blaze
function) of the template is unknown. However,
we are only interested in correcting relative
changes compared to the template, so a low
order polynomial correction should be
sufficient given a stabilized instrument such
as HARPS. Section \ref{sec:blaze}
investigates the optimal degree required for
such polynomial.

All the analysis described above is done in
barycentric wavelengths. The correction from
observed to barycentric wavelengths is also
implemented using the aforementioned simplistic
recipe for the Doppler factor using the projected
radial velocity of the observer as computed by
\citet{stumpff:1980} (also provided in header of
the HARPS DRS files). Therefore, as a first guess,
$\alpha_v$ is initialized to 1. Before the
least-squares adjustment begins, $\alpha_v$ is
initialized by finding the approximate $\chi^2$
minimum using \ref{eq:chi2} in a bisection
algorithm with $\alpha_v$ as the independent
variable. The obtained RV from this bisection is
typically within 10 \ms of the final RVs,
minimizing the required number of computationally
expensive least-squares iterations.

Note that the simplistic definition of the Doppler
factor we are using is only accurate to first
order on $v/c$. Since we are only interested in
differential RV measurements, one can show that
the second order terms ($\sim v^2/c^2$ but also
gravitational terms proportional to $G$) are
mostly constant and that this expression for
$\alpha_v$ is good at the few tens of c\ms. While
higher order terms should be certainly taken into
account, such refinement is useless unless a full
revision of the Barycentric correction model is
implemented. Critical parts in this model that
need to be revised are: 1) use of an up-to-date
model for Earth rotation, 2) use of the most
recent version of the ephemeris for Earth \citep[e.g.,
DE405 or newer, ][]{standish:1998} and 3) inclusion 
of all the relativistic corrections to the first
post-Newtonian order (gravitational and
kinematic). Even though such model might improve
some of the RVs presented here, further work is
required to produce a functional and reliable code. A
tool to provide Barycentric Doppler corrections to
$1$ c\ms level is in development and we plan to
present it in a future publication.

\subsubsection{Convergence criteria}

Since the Doppler factor is the only strongly
non-linear parameter, the convergence criteria is
established as the iteration requiring a $\delta
\alpha_v \times c$ smaller than  $1$ c\ms. Thanks
to the simplicity of the model, only 3 to 6
iterations are typically required to reach
convergence.

The typical nominal uncertainty of an echelle
aperture with a SNR of 100 is between 2 to 15
\ms, strongly depending on spectral type and
aperture under consideration. Examples of
template matched observations for a G8.5V star
(Tau Ceti) and an M4 dwarf (GJ 699) are shown in
Figures \ref{fig:spectrumfitG8} and
\ref{fig:spectrumfitM4}.

\subsection{Obtaining the final RV measurements}\
\label{sec:finalrv}

A final refinement is useful to produce the
highest quality RV. Because the template has
uncertainties, there can be ambiguous radial
velocity zero-point offsets associated to each
aperture. Also, some apertures might have extra
uncertainty due to instrumental effects not
accounted in the formal uncertainties (extraction
issues, problems with the wavelength solution,
etc.). With the purpose of computing a more
realistic weighted mean to each epoch, the
zero--point and nominal uncertainty of each order
can be empirically reassessed as follows.

As a first step, the weighted mean $\mu_e$ of each
epoch $e$ is computed as

\begin{eqnarray}
\mu_e &=& \frac{1}{Z}\sum_{a=0}^{N_a}\frac{v_{e,a}}{\sigma_{e,a}^2}\, ,\\
Z &=& \sum_{i=0}^{N_a} \frac{1}{\sigma_{e,a}^2}\,.
\end{eqnarray}

\noindent where the subindex $a$ runs over all the
$N_a$ apertures being analyzed, $v_{e,a}$ is the
RV measurement of the $a$-th aperture at epoch
$e$, and $w_{e,a}$ is the corresponding formal
uncertainty in the RV ($\sigma_{\alpha_v}$ as
obtained from equation \ref{eq:formalerrors}
multiplied by the speed of light). Z is the sum of
all the weights and normalizes the weighted mean.

Let us now concentrate in one aperture, say
aperture 15. On each epoch $e$, we compute the
difference between the epoch weighted mean $\mu_e$
and $v_{e,15}$. The weighted average of this
difference over all the epochs is the zero--point
of order 15, which we call $v^{(0)}_{15}$. The
weighted standard deviation of these differences
over all the epochs is the new nominal uncertainty
$\sigma^\prime_{e,15}$. This process is then
repeated for all the orders to obtain the
corresponding $v^{(0)}_{i}$ and
$\sigma^\prime_{e,i}$. After all the weights and
zero-points are obtained, the final radial
velocity measurement and its precision on a given
epoch is given by
\begin{eqnarray}
RV_e &=& \frac{1}{Z^\prime}
\sum_{i=0}^{N_o}\frac{v_{e,i}-v^{(0)}_{i}}{\sigma^{\prime 2}_{e,i}}\, ,\\
\sigma_e^2 &=& \frac{1}{Z^\prime(N_o-1)}
\sum_{i=0}^{N_o}\frac{(v_{e,i}-v^{(0)}_{i})^2}{\sigma^{\prime 2}_{e,i}}\, ,\\
Z^\prime &=& \sum_{i=0}^{N_o} \frac{1}{\sigma^{\prime 2}_{e,i}}\,.
\end{eqnarray}
\noindent Note that, because $v^{(0)}_{i}$ and
$\sigma^\prime_{e,i}$ are computed with
respect to the initial guess of the epoch mean
value, the zero--point correction does not
remove offsets shared by all the apertures
(i.e. Keplerian signals are not removed). In
case the user is interested in the RV
measurements without this re-normalization of
the weights, the original $\mu_e$ and the
individual RV measurements of each order are
also provided as a data product by
HARPS-TERRA.

As mentioned in the introduction, these
algorithms and all the code necessary to
extract the HARPS spectra from the ESO reduced
data--products are included in the HARPS-TERRA
software. HARPS-TERRA has been designed to achieve
the highest precision but while being computationally
efficient. It takes between 1 and 5 seconds to
process one full spectrum on a Linux operating
system running on a 2.0 GHz processor. The
construction of a high SNR template (see next
Section) takes around 5 minutes when using 200
spectra. A data set consisting on 100 spectra
can be fully processed in less than 10
minutes.

\subsection{Construction of the template}\label{sec:template}

In order to obtain maximal precision, one would
like a template with the highest possible SNR.
This can be achieved by carefully coadding all the
available spectra. Still, some precautions must be
taken in the construction of such template. This
is an outline of the process we use to generate
them.

In a first pass over all the spectra, the RV and
flux normalization coefficients are obtained with
respect to a preliminary template (highest SNR
observation). This preliminary RV measurement and
the heliocentric motion of the observer are used
to obtain the barycentric wavelengths of each
observed spectrum at each epoch.

Spectra obtained in different epochs are not
sampled at the same barycentric wavelengths and
cannot be co-added without some kind of
interpolation. To solve this, the preliminary
template is used as a reference to generate a grid
of regularly spaced reference wavelengths. The
number of re-sampled reference wavelengths is 4
times the number of original pixels in each
aperture of the preliminary template (this is
$4096\times 4 = 16384$). Each observed spectrum
can then be interpolated on such reference
wavelengths using a cubic spline. Finally, the
flux normalization is applied and the template
value at each re-sampled wavelength is computed as
a 3-$\sigma$ clipped mean over all the epochs.


The flux normalization polynomial is only applied
if the average SNR of a given aperture is higher
than 5. SNR lower than 5 are not rare on the bluer
apertures of M dwarfs and the corresponding
normalization polynomials are very unreliable.
Because the blaze variability is of the order of
1\%, it is safer to simply apply a scale factor
and match the average flux of the observation to
the average flux of the preliminary template.

Telluric features must also be removed from the
coadding. As in the RV measurement, the flux
measurements coincident with telluric features
deeper than 1\% are masked out. Let us note that,
if a star has been observed in different phases of
the Earth motion around the Sun, this process
allows to generate a telluric free spectrum on
regions with mild telluric contamination.
When a significant number of spectra is
available (N$>$20), the result of this
coadding can be very spectacular, specially on M
dwarfs whose spectrum has almost no continuum
and it is very hard to distinguish pure noise
from real features (see Figure
\ref{fig:spectrumfitM4} as an example).

\section{Performance}\label{sec:parameters}

Even though the described algorithms are
relatively simple, the extraction of precise
radial velocities still depends on a number of
parameters that have to be tuned by hand. The
two parameters we investigate here are; (1)
the optimal degree for the flux normalization
polynomial, and (2) bluest echelle aperture
to be used (e.g., M dwarfs can have 10 to 100
times more flux on the red than on the blue
and low SNR spectra are typically more
affected by systematic noise). HARPS-TERRA has
been designed to be very flexible on all such
parameters but a general procedure to produce
optimal results with minimal human
intervention is still desirable. To illustrate
the effect of these two parameters we use
observations of Tau Ceti (GJ 71, HD 10700, a
very stable G8.5V dwarf), HD 85512 (GJ 370, a
quiet K6V dwarf with a very low amplitude
candidate planet) and Barnard star (GJ 699, a
relatively quiet halo M4 dwarf). These three
stars have abundant data in the ESO archive
and bracket the highest priority targets for
the search of very low mass companions (G, K
and M dwarfs). The direct comparison of the
final RVs obtained with CCF and HARPS-TERRA on
a larger sample is given in Section
\ref{sec:comparison}. All the stars we discuss
here are nearby and show a significant linear
trend due to perspective acceleration
\citep{zechmeister:2009}. In all that follows,
such perspective acceleration has been
subtracted from the measured RVs.

\begin{deluxetable}{crrrrrrrrr}  
\tabletypesize{\scriptsize}
\tablecolumns{9}

\tablecaption{Relevant parameters to this work
of each star. A more detailed description can
be found in the SIMBAD database and references
therein. All quantities are given to the last
significant digit.}

\tablehead{
  \colhead{Parameter} &
  \colhead{GJ 676A} &
  \colhead{Tau Ceti} &
  \colhead{HD 85512} &
  \colhead{Barnard's} &
  \colhead{Kapteyn's} &
  \colhead{Proxima} &
  \colhead{$\epsilon$ Eri} &
  \colhead{HD 69830} &
}
\startdata
$\mu^*_{\rm R.A.}$
$[$mas yr$^{-1}]$
\tablenotemark{a}       & -260    & -1721 &  461   & -798  &  6505  & -3776 & -975  &  279        \\
$\mu_{\rm Dec}$
[mas yr$^{-1}$]
\tablenotemark{a}       & -184    &  854  &  -472  & 10328 & -5731  &   766 &  19   &  -987       \\
Parallax [mas]
\tablenotemark{a}       & 61      & 274	  &  90    & 548   & 256    &   772 &  311  &  80          \\
V\tablenotemark{b}      & 9.58    & 3.5	  &  7.651 & 9.51  &  8.85  & 11.05 &  3.73 &  5.95          \\
K \tablenotemark{b}     & 5.82    & 1.79  &  --    & 4.52  &  5.05  & 4.38  &  1.78 &  4.16         \\
Sp. type
\tablenotemark{b}       & M0V     & G8.5V &  K6V   & M4V   &  M1V   & M6V   &  K3V  &  G8V           \\
Mass [M$_\sun$]
\tablenotemark{c}       & 0.71    & 0.78  &  0.69  & 0.16  &  0.27  & 0.12  &  0.82 &  0.86          \\
\enddata

\tablenotetext{a}{Proper motions and
parallaxes from HIPPARCOS
\citep{hipparcos:2007} are required to
subtract the perspective acceleration effect
(see text). $\mu^*_{\rm R.A.}$ corresponds to 
$\mu_{R.A.} \cos \delta$ and is the proper motion
in the direction of increasing R.A. in a local tangent 
plane defined on the star's nominal coordinates at 
the catalog reference epoch. $\mu_{R.A.}$ 
is the obsolete coordinate dependent definition
of the secular change in R.A., which is
singular at the celestial poles and has not direct
physical interpretation.
}

\tablenotetext{b}{Unless noted in the
manuscript, V and K photometry and nominal
spectral type are obtained from the SIMBAD
database.}

\tablenotetext{c}{Stellar masses for M dwarfs
have been derived from \citet{delfosse:2000}
using absolute K magnitudes. These masses
should be accurate at the 5-10\% level. Masses
of the K and G dwarfs are obtained from
various references (see section on each
star).}

\label{tab:stars}
\end{deluxetable}

\textbf{Tau Ceti}. For this experiment we use 84
spectra of the 4000+ available in the HARPS-ESO
archive. Since it is a very quiet and stable star,
Tau Ceti had been used as a standard star by
several HARPS programs. Not all such programs aim
to achieve the highest RV precision. This results
in a very heterogeneous sample of public
observations with varying exposure times and a
wide range of SNR. The SNR of some of the
available spectra is so high ($>500$) that
saturation or extraction problems seem to be
seriously affecting the reddest orders. For a
consistent analysis, we prepare a sub-sample from
the \textit{HARPS high precision survey for
exoplanets in the southern hemisphere}
\citep[e.g.][]{lovis:2006,mayor:2009}. This subset
also belongs to the sample presented in
\citep{pepe:2011} showing an RMS of 0.92 \ms over
a time-span of 5 years. \citet{pepe:2011} noted
that, to achieve such precision, several spectra
in a given night had to be averaged to mitigate
the effects of stellar pulsation and granulation.
Because this star is one of the 10 higher priority
targets in the \textit{HARPS-GTO survey for
Earth-like planets}, only the observations taken
before 2009 are publicly available. In any case,
we only need a statistically significant sample
with consistently measured RVs using the CCF
method to be used as a reference. Still, to work
with a more manageable sample, we only use the
first observation of each night. The final sample
contains exposure times between 30 and 150 seconds
and a typical SNR in the range between 100 to 250
at 6000 \AA. The CCF RVs for this sub-sample show
an RMS of 1.53 \ms. The RMS from HARPS-TERRA using all
the apertures is 1.52 \ms. Since we are only
using one spectra per night; this sub-sample
shows, as expected, a larger RMS that the nightly
averages used in \citet{pepe:2011}.

\textbf{HD 85512} (GJ 370), is a quiet K6V dwarf
with a very low amplitude candidate planet
\citep{pepe:2011} that could support liquid water
if it were significantly covered by clouds
\citet{kaltenegger:2011}. It is also a target of
the current HARPS-GTO program to search for low
mass companions, so only the first 5 months of
observations (Dec 2008 to March 2009) are publicly
available. The typical integration times are
between 400 and 600 seconds and the SNR at 6000
\AA\, ranges from 100 to 250. For our purpose
here, it is only relevant that the CCF
measurements show an RMS as low as 1.10 m s$^{-1}$
using a K5 binary mask (a very stable star
indeed). The HARPS-TERRA RVs from the same 122
observations show an RMS of $\sim 1.0$ m s$^{-1}$,
which already represents a significant
improvement.  For comparison, the proposed planet
candidate has an RV semi-amplitude of 0.8 m
s$^{-1}$ and the detection is based on 250+
measurements in \citet{pepe:2011}. A quadratic
trend (also present in the Ca H+K S-index activity
index) had to be removed to cleanly detect the
candidate in \citet{pepe:2011}. No trend nor planet
signals were subtracted to the RVs discussed here.

\textbf{Barnard's Star} (GJ 699) is the star with
the highest proper motion and the second closest
star system to the sun. It is an M4 dwarf with
halo kinematics (total velocity with respect to
the Sun is $\sim$ 150 km s$^{-1}$) and it is known
to be slightly metal poor. Even though it is a low
mass star and relatively faint in absolute terms,
its proximity to the Sun (1.82 pc) allows to
obtain a typical SNR between 50 and 80 at 6000
\AA\,in 900 sec integrations. 22 spectra are
available in the HARPS-ESO archive over a
time-span of one year (April 2007 to May 2008).
Even though it is classified as active (V2500
Oph), only occasional flares have been reported on
it. Barnard star was observed within the
\textit{ESO-UVES search of low mass companions
around M dwarfs} \citep{zechmeister:2009}. Those
measurements demonstrated its radial velocity
stability down to 2.5 \ms using the Iodine cell
technique. They also found that a possible RV
wobble with a 40 days period was significantly
correlated with the strength of its H$_\alpha$
emission. Therefore, we \textit{a priori} expected
that the RV measurements obtained with HARPS would
also show some activity induced jitter. The CCF
RVs as extracted from the archive are obtained
using an M2 binary mask and show an RMS of 1.54
\ms. The RVs from the HARPS-TERRA RV measurements
using the full spectrum have an RMS of $1.23$ m
s$^{-1}$, which again represents a significant
improvement. A quick-look analysis of the RVs does
not show evidence of the previously reported
periodicity at 40 days, but the number of
observations is too low and the time-sampling is
too sparse to rule out signals in this period
domain.

\textbf{Two more stable M dwarfs}. In addition
to Barnard's star, we will briefly use HARPS
RV measurements of two other M dwarfs :
Proxima (GJ 551, M6Ve) and Kapteyn's star (GJ
191, M1V subdwarf). The data set on Proxima
consists on 27 radial velocities taken between
May 2005 and February 2009. The CCF RMS is
2.29 m s$^{-1}$ and recurrent flaring events
can be detected in more than one epoch. The
data set on GJ 191 contains 30 spectra taken
between December 2003 and February 2009. The
CCF RMS is 2.45 m s$^{-1}$. These two stars
will be used to illustrate some features that
seem to be common on M dwarfs other than
Barnard's.

\subsection{Flux normalization polynomial}\label{sec:blaze}

Because RV measurements rely on the slopes of
strong spectral features, the flux normalization
correction is a key element to achieve the highest
possible RV precision. For example, much better
precision is obtained from deep-sharp lines than
broad-shallow ones. In a perfectly calibrated
instrument, the blaze function should be constant
over time. However, several
instrumental/observational effects can cause
variability of the effective blaze function. For
example, due to atmospheric differential
refraction, the photocenter of a star at the
entrance fiber will depend on wavelength causing
wavelength dependent flux loses as a function of
the airmass. Also, bluer wavelengths are more
efficiently dispersed by the atmosphere adding
additional airmass dependent variability. At the
end of the day, the combination of several effects
causes time variable blaze function shapes at the
level of a few \% which (as shown below) can
severely perturb the RV measurements.

\begin{figure*}[tb]
\centering
   \includegraphics[width=5.5in,clip]{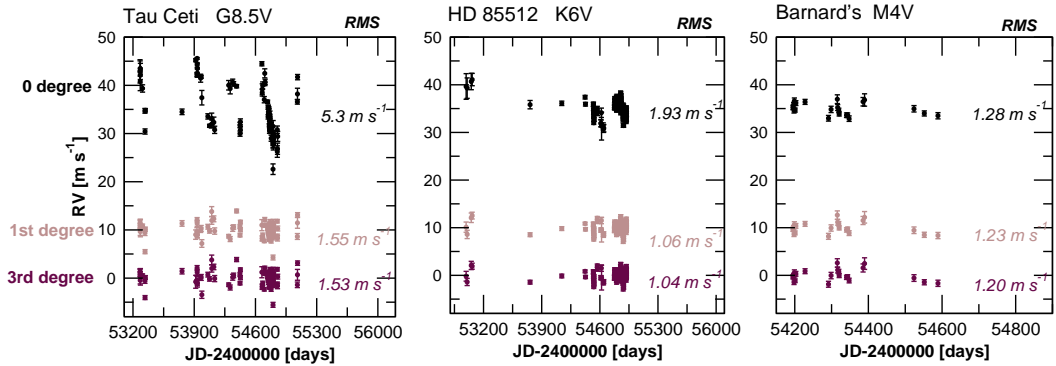}

\caption{Radial velocity measurements as a
function of time obtained on the three test
stars using different flux normalization
polynomials. Unless corrected, the flux
variability strongly perturbs the RV
measurements on the G star (left) as can be
seen in the obtained RMS when only an overall
normalization factor is applied (order 0 in
black). The effect of the flux variability is
still significant for the K dwarf and has very
little effect on the M dwarf. An order 1
polynomial (blue) provides a significant
improvement on both the G and the K dwarfs and
an order 3 polynomial (red) seems to reach a
good compromise between computing efficiency
and precision on all spectral types. RV
offset has been added to the measurements
of the 0 and 1 degree polynomials to
improve visualization.} \label{fig:rmsvsblaze}

\end{figure*}

Because we are interested in relative
measurements, we only need to correct for the
differential variations of the blaze as compared
to the template. In figure \ref{fig:rmsvsblaze},
we show the RV measurements of our three test
stars when using different flux normalization
polynomials. When only a constant
normalization is applied (0 degree polynomial),
the scatter in the radial velocities of the G and
the K dwarfs is very large and has a clearly
systematic behavior. The RV precision dramatically
improves when using a linear flux correction
(first degree polynomial) on both stars. No
significant improvement in precision is obtained
by using polynomial degrees higher than 3. On the
other hand, the RV measurements on the M dwarf are
much less sensitive to the degree of the polynomial
used. The improvement on the precision is modest
but still significant with a linear correction,
and optimal results are also obtained when a cubic
polynomial is applied.

The sensitivity to the flux normalization
correction as a function of the spectral type was
expected. The spectra of typical G and K dwarfs
consist on well isolated sharp lines against a
smooth continuum. If not properly corrected, the
changes in the slopes induced by the blaze
variability on the continuum contain spurious
Doppler information that strongly perturbs the
Doppler measurement. On the other hand, the
spectrum of an M dwarf is dominated by heavily
blended molecular absorption bands which strongly
dominate over the shifts induced by blaze
variability. Given that high degree
flux corrections are computationally expensive,
we set the nominal blaze correction to a
cubic polynomial.

\subsection{Most useful echelle apertures}
\label{sec:apertures}

The final uncertainty in the radial velocity
measurement depends on many instrumental effects
in addition to the formal statistical
uncertainties and the stellar spectrum. K and,
specially, M dwarfs have significantly less flux
in the blue than in the red. If noise in bluer
wavelengths were purely random, these apertures
would be properly down-weighted through the formal
uncertainties and the final measurement would be
unaffected. However low SNR spectra can be more
sensitive to instrumental systematic effects, so
using arbitrarily low SNR data can be
counterproductive. As an example, the M2 binary
masks used by the HARPS DRS, only uses apertures
redder than the 22nd one ($\lambda > 4400 $\AA).
Also, given that all the stars are active at some
level and that activity should affect the measured
RV differently at different wavelengths
\citep{reiners:2010}, one could expect some
apertures to provide more reliable measurements
than others. As an example, star spots are known
to have higher contrast at bluer wavelengths, so
one would expect stronger RV stellar jitter on the
blue. The wavelength dependence of the stellar
jitter has been exploited in the past to rule out
possible companions around young stars using
complementary RV measurements in the near
infrared, e.g., see the RV measurements obtained
in the H band by
\citet{huelamo:2008,figueira:2010} on TW Hya using
the CRIRES spectrograph at the ESO/VLT. This
section is devoted to develop a strategy to
determine the wavelength dependence of the RV
precision and define the bluest aperture to be
used for each spectral type.

\begin{figure*}[tb]
   \centering
   \includegraphics[width=140mm,clip]{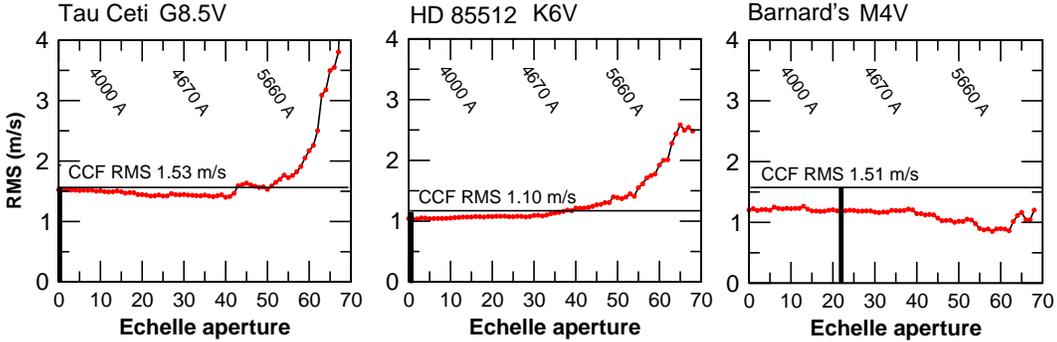}

\caption{Radial velocity RMS as a function of
the bluer aperture used. The RMS derived from
the public CCF measurements is indicated as a
horizontal line. The bluer aperture used in
the CCF analysis is marked as a vertical thick
line. While the RMS of the K6V dwarf decreases
as bluer apertures are added, neither the M
dwarf nor the G dwarf show the same behavior.
The origin of this RMS excess in the blue is
still unknown. The fact that it only shows on
some stars very stable stars, indicates that
is likely due to stellar induced jitter rather
than an instrumental effect.}

\label{fig:rmsvsorder}

\end{figure*}

Given that the typical formal uncertainty of a
given aperture is never better than 2-3 \ms, it
is hard to distinguish random noise from
systematic effects (intrinsic to the star or
instrumental) when looking at the RV measurements
of single echelle apertures. Instead, we use the following
procedure to assess the dependence of the
precision with the bluest aperture used. First,
we extract the radial velocities on all the
apertures. Then we measure RVs on all the epochs
as described in Section \ref{sec:finalrv} but only
using the apertures between 71 (reddest available
one) to some bluer one (say aperture 68), and
obtain the RMS of these time series. We then
repeat this process adding one aperture at a time
and plot the obtained RMS as a function of the
bluest aperture used. This is illustrated on
Figure \ref{fig:rmsvsorder} on our three test
stars. On the right hand-side of each panel, the
RMS is expected to be higher because less
apertures are used. However, if the noise were
purely random, we would expect a reduced RMS as
more apertures are added to the blue. The K6V
dwarf nicely follows this behavior.

\begin{figure*}[tb]
   \centering
   \includegraphics[width=4in,clip]{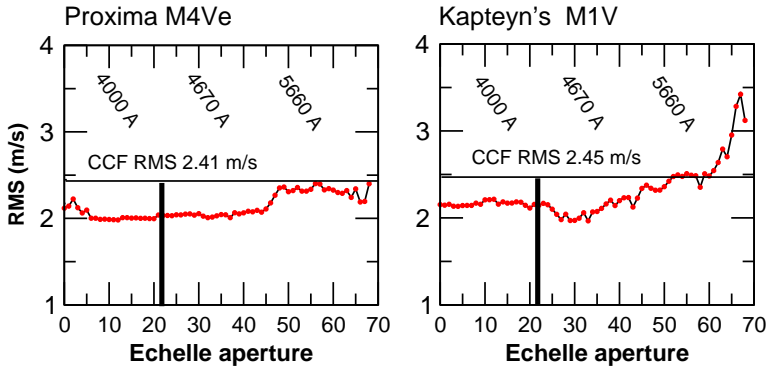}

\caption{RMS as a function of the bluer aperture
used on Proxima and Kapteyn's star (GJ 191)
showing a similar behavior as Barnard star. This
is, the RMS starts decreasing as bluer wavelengths
are added until it starts increasing again (GJ
191) or does not change significantly (Proxima).
For Proxima, the SNR below aperture 10 is so low
(SNR $<5$) that adding such apertures to the
analysis only adds systematic noise to the RV
measurement. As a general rule, the cut at
aperture 22 used by the HARPS DRS seems a
reasonable assumption.} \label{fig:rmsvsorderM}

\end{figure*}

However, the M4V star behaves quite differently.
The RMS reaches a minimum (81 c\ms) at aperture
58 when only the 14 reddest apertures are used.
Then, it starts increasing as more apertures are
included towards the blue($\sim$ 1.2 \ms at HARPS
aperture 0). The typical SNR is still very
significant on apertures bluer than 58 (SNR $>$
30) so instrumental systematic effects are not
expected to be important at this level. This
combined with the fact that the K6V star does not
show such an excess, suggests that a significant
fraction of this extra noise is not instrumental
but intrinsic to the star. For this star in
particular, the 58-th aperture ($\lambda\sim$
6000\AA) seems to be the middle point where the
activity induced variability and the
photon+instrumental noise equally contribute to
the error budget. This behavior is also present
in other \textit{stable} M dwarfs. For example, we
show the same diagrams for Proxima (M6V) and
Kapteyn's star (M1.5V, GJ 191) in Figure
\ref{fig:rmsvsorderM}. Because both stars are
fainter than Barnard's, one has to use bluer
apertures to reach the point where the
instrumental noise meets this wavelength dependent
jitter. In an ideal situation the bluest
aperture should be selected for each star, but if
this effect is really related to stellar activity,
this sweet spot is difficult to predict a priori.
This behavior is under investigation and hints
towards new ways of assessing if a signal could be
activity induced. Also, the wavelength dependence of
the RMS illustrates the importance of moving to
redder wavelengths to efficiently searh for low mass companions
around cool stars \citep{reiners:2010}. Figures
\ref{fig:rmsvsorder} and \ref{fig:rmsvsorderM}
also show the RMS obtained from the CCF
measurements(horizontal line) illustrating that
the template matching technique works better even
without \textit{cherry picking} the bluer
aperture to be used. As mentioned before, the CCF
M2 binary masks used by the HARPS DRS do not
consider apertures bluer than the 22nd. Given
that aperture 22 also seems a reasonable
choice for HARPS-TERRA, we will also use it
as the default bluer aperture to be used when
extracting RVs from M dwarfs.

Tau Ceti is known to be one of the most RV stable
G dwarfs. A bit unexpectedly, we find that the RMS
as a function of the bluer aperture also shows a
minimum of 1.36 \ms at aperture 40, and then it
starts increasing again towards the blue (1.52 \ms
at aperture 0). Given that the SNR is still high
($\sim$ 40 at order 0), again this should not happen in
a perfectly stable star. As for the M dwarf, this
indicates that some scatter is due to stellar
activity and has a chromatic component (e.g.
stellar pulsation should be achromatic). The RMS
obtained at the sweet-spot (order 40th) can be used
to estimate the magnitude of this chromatic noise
as $1.36/\sqrt{2} \sim 0.96$ \ms, which is quite
significant. For comparison, an Earth-mass in the
habitable zone on Tau Ceti would have an RV
amplitude of 10--20 c\ms. Given that it is one of
the most quiet Sun-like stars being surveyed for
rocky planets, this issue requires a more detailed
analysis which is beyond the scope of this paper.
Given the increased precision achievable by
selecting the sweet spot aperture, a reprocessing
of the full data set in \citet{pepe:2011} with
HARPS-TERRA could lead to a significant increase
in the overall precision and enhanced sensitivity
to lower mass companions.

In conclusion to this section, we find that a
cubic polynomial is sufficient to obtain an
optimal flux normalization correction. The maximal
precision on quiet stars (such as HD
85512) is obtained using the whole spectrum.
However, in several stars a moderate improvement
in the overall precision could be achieved by
finding the sweet spot between random noise
(instrumental+photon) and the wavelength dependent
jitter. The determination of this sweet-spot
requires a better understanding of its physical
origin and will not be further discussed here. To
make a fair and consistent comparison, all the
apertures will be used when analyzing spectra of
G and K dwarfs. For M dwarfs, only apertures
redder than aperture 22 ($\sim 4400$ \AA) will be
used. This bluest aperture used together with the
cubic flux normalization polynomial discussed in
Section \ref{sec:blaze} define the
\textit{standard setup} of HARPS-TERRA when
analyzing spectra from G, K and M dwarf stars.

\subsection{Ca II H+K activity indicator}
\label{sec:svalue}

Pseudo-radial velocity variations can be caused by
stellar activity. As mentioned before, it is
suspected that some apparent RV offsets are
closely related to the magnetic activity of the
star and related surface features such as spots.
Also, enhanced magnetic activity can cause local
or global changes in the convection patterns of
the stellar surfaces. When convection is enhanced,
hotter and bluer material emerging from the
convection cells (e.g. stellar granulation) causes
apparent blue shifts to the integrated stellar
spectrum. For one reason or another, one could
expect apparent radial velocity jitter whenever
the magnetic field of the star experiences
changes. A detailed discussion on the topic can be
found in \citet{lovis:2011} and references
therein. \citet{lovis:2011} also provides the
recipes to compute one of the most commonly used
activity indicators, the Mount Wilson S-index,
using HARPS data. This index measures the relative
flux of the Ca II H and K lines in emission
($\lambda_H = 3933.664\AA$ and $\lambda_K =
3968.470$) compared to a local continuum. These
lines in emission are formed in the hot plasma of
the chromospheres of stars and, as a consequence,
their intensity varies with the strength of the
stellar magnetic field. We incorporated the
automatic measurement of the S-index as one of the
outputs of HARPS-TERRA. To obtain the correct flux
estimates, it is necessary to know the absolute
heliocentric radial velocity of the star at few
hundred \ms accuracy. Since HARPS-TERRA cannot
provide this information, we use the median of the
heliocentric CCF RV measurements to estimate the
barycentric wavelengths of the lines on all the
epochs. Because the H and K lines and the
continuum bands defined in \citet{lovis:2011}
appear in different echelle apertures (aperture
5 and 6), we use the re-sampled, blaze corrected
full spectrum also provided by the HARPS DRS. The
full spectrum is then interpolated using a cubic
spline and the indices are computed on a regularly
sampled grid of 0.01 \AA conveniently Doppler
shifted to match the heliocentric radial velocity
of the star.

\begin{figure*}[tb]
   \centering
   \includegraphics[width=140mm,clip]{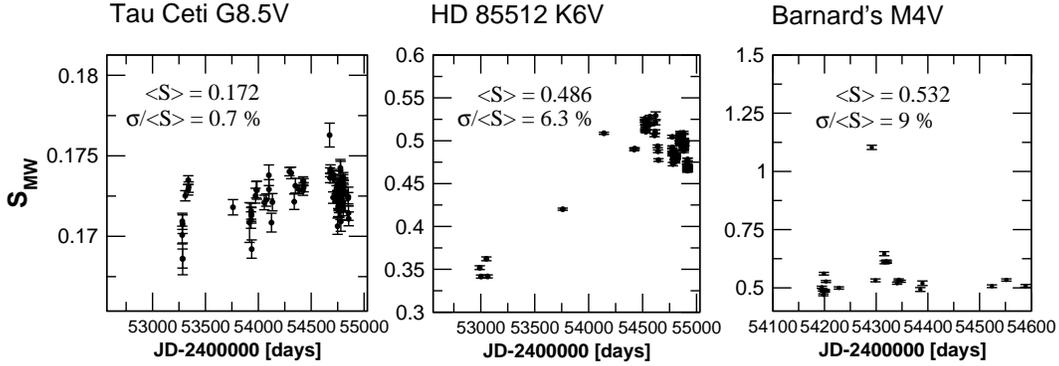}

\caption{S-Index time series on Tau Ceti, HD
88512 and Barnard's star. The mean value
$\left<S\right>$ and the relative variability of
the index $\sigma/\left<S\right>$ are given for
each star. One of the epochs for Barnard star
shows an S value two times higher than usual
which is indicative of a flaring event occurring
during the observation. That point was excluded
from the computation of the mean S value and its
relative variability. The error bars represent
the photon noise in the measurement of S.}
\label{fig:sindex}

\end{figure*}

To validate this procedure, we obtained
time-series of the S-index on published HARPS
stars obtaining perfect agreement in all
cases. The S-index measurements on Tau Ceti
and HD 85512 are illustrated in Figure
\ref{fig:sindex} and show the same behavior
reported by \citet{pepe:2011}. No coherent
variability is observed on Tau Ceti (relative
scatter is $0.7\%$), and long term variability
is observed on HD 85521. The S-index as a
function of time is also plotted for Barnard's
star. Barnard's (M4V) shows a similar amount
of relative variability as the K dwarf except
for a mild flaring event that doubles the Ca
II H+K flux in one of the epochs. In case a
tentative signal is found, the time series of
the S-index will be used to investigate
possible correlation of the RVs with magnetic
activity. An example of this is illustrated in
section \ref{sec:HD69830}, where we find that
a promising long period quadratic trend in the
RV is correlated with an almost identic trend
in the S-index. A popular form of the same
activity indicator is the so--called
R$^\prime_HK$ index. R$^\prime_HK$ represents
the flux in emission of the Ca H+K lines over
the total stellar spectrum and requires some
assumptions on the spectral energy
distribution of the star. Still, R$^\prime_HK$ is obtained
as a linear relation from S and, as a
consequence, it will experience the same time
variability. Since the computation of
R$^\prime_HK$ requires additional information
about the star \citep[e.g. effective
temperature and metallicity][]{lovis:2011} and
does not provide extra information about the
time variability, R$^\prime_HK$
is not provided by HARPS-TERRA.

The HARPS DRS also provides three other activity
indicators derived from the shape of the
Cross-Correlation Function. The first one is the
so-called Bisector Span of the CCF (BIS). BIS is a
measure of the asymmetry of the average spectral
line and should correlate with the RV if the
observed variability is caused by spots or plages
rotating with the star \citep{queloz:2001}. The
second index is the full-width-half-maximum (FWHM)
of the CCF and is a measure of the width on the
mean spectral line. The variability of the FWHM is
thought to be a direct consequence of changes in
the convective patterns on the surface of a star;
effectively changing the shapes of the integrated
line profiles but could also depend on other
physical processes related to the stellar magnetic
field. The third index is the
contrast of the CCF (CONTRAST), which is sensitive
to the changes in the depth of the average
spectral line profiles. These three numbers
provide important diagnostics to distinguish
genuine Doppler signals from activity induced
periodicities. Since they are already computed by
the HARPS DRS, they  will not be further discussed
here. Equivalent activity indicators optimally
designed for the template matching technique are
in development and will be given in a future
publication.

\section{Comparison : CCF vs template matching}
\label{sec:comparison}

Comparing actual radial velocities is more
informative than discussing which method (CCF
or template matching) works better on
theoretical grounds. This section is devoted
to illustrate the performance of the template
matching approach compared to the pipelined
HARPS CCF measured RVs. We discuss a few
remarkable systems in terms of reported planet
abundance but also their reported RV
stability. We show that, while the performance
on G and K stars is similar using both
techniques, template matching works
significantly better on M dwarfs and
moderately active stars. Given the relative
youth of the HARPS-TERRA code compared to the
many years of refinement of the HARPS-CCF
algorithms, we can only expect further
improved precision in the future. In addition
to alternative RV measurements, the template
matching approach should allow us to perform a
new set of diagnostics such as the
determination of the bluest aperture to be
used for an optimal RV extraction already
discussed in Section \ref{sec:apertures}.

Whenever signals are present and orbital fits
are required, we use the SYSTEMIC interface
\citep{meschiari:2010} as provided in August
2010\footnote{http://oklo.org} to obtain the
orbital parameters and their uncertainties.
SYSTEMIC allows the interactive adjustment of
multi-planetary systems and is able to generate
a large variety of data products and figures.
When periodograms are required to illustrate
the detection of a signal, we use a custom
made version of a least-squares periodogram. A
sinusoidal signal is adjusted to a list of
10$^4$ test periods between 1.1 and 10 000
days. The F-ratio statistic of the
least-squares solution is then plotted against
the period so the higher peak represents the
most likely periodicity in the data. These
periodograms are based on the definitions
given in \citet{cumming:2004}, and they are
formally equivalent to the so-called
generalized Lomb-Scargle periodograms (GLS)
described by \citet{zechmeister:2009b}. The
CCF and HARPS-TERRA RV measurements used in
this section are given Appendix
\ref{sec:rvdata}.

\subsection{Proxima, a flaring M4V}

Proxima Centauri is the nearest stellar neighbor
to the Sun and, therefore, has a special interest
among the planet hunting community and the public
in general. So far, there have not been any
serious claim of a companion around this star. It
is the common proper motion pair to the $\alpha$
Centauri binary, two stars with masses similar to
the sun in a long period orbit. Proxima Cen has been
intensively monitored by the ESO/UVES planet
search program at a median precision of 3 \ms
using the Iodine cell technique and no significant
periodicity was detected \citep{endl:2008}.
Between 2004 and 2009, it has been observed 27
times with HARPS on different programs.

\begin{deluxetable}{lrrr}  
\tablecolumns{4}
\tabletypesize{\scriptsize}
\tablecaption{Overview of the RV measurements on the three M dwarfs without planets
discussed in the text.}
\tablehead{
  \colhead{Parameter} &
  \colhead{Proxima} &
  \colhead{Barnard's} &
  \colhead{Kapteyn's}\\
}
\startdata
Mean SNR at 6100 \AA\            & 39	      & 120      & 85         \\
Number of RV                     & 27         & 30       & 22         \\
First observation                & May 2004   & Dec 2003 & April 2007 \\
Last public observation          & Feb 2009   & Feb 2009 & May 2008   \\
RMS$_{\rm CCF}$ [\ms]            & 2.37       & 2.45     & 1.51       \\
RMS$_{\rm TERRA}$ [\ms]          & 2.05	      & 2.13     & 1.19       \\
Extra noise in CCF [\ms]         & +1.18      & +1.21    & +0.92      \\
\\
Optimal bluer aperture          & 37         & 29       & 56         \\
RMS at optimal aperture [\ms]   & 1.94       & 1.95     & 0.82       \\
\enddata
\label{tab:mdwarfs_overview}
\end{deluxetable}

\begin{figure}[tb]
\centering
\includegraphics[width=3.0in,clip]{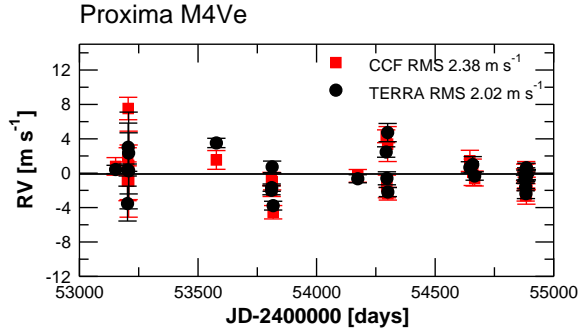}
\caption{Measured radial velocities on Proxima
using HARPS-Terra (black) compared to the ones
obtained with the CCF method(red)}
\label{fig:Proxima_rvs}

\end{figure}

Visual inspection of Proxima's spectrum strongly
suggests it is an active star as several activity
indicators in the HARPS wavelength range are in
strong emission ($H_\alpha$, NaD, Mg, Ca, etc).
Perhaps for this reason and because it is
relatively faint, the star has not been as
intensively monitored as other earlier type
nearby M dwarfs such as GJ
581\citep{forveille:2011b} or GJ 876
\citep{rivera:2010}. The HARPS-TERRA RVs have
been obtained using the standard setup for M
dwarfs. Figure \ref{fig:Proxima_rvs} shows the
CCF and HARPS-TERRA measurements as a function of
time. In July 16 2004, Proxima experienced an
energetic event (probably a flare) causing all
the activity indicators to go to strong emission.
The flare event happened on the second epoch and
the emission lines (such as H$_\alpha$) show
intensities 2 to 10 times stronger
than their quiescent state (S-index goes from a
median value of 8.7 to 58). Still, neither the
CCF nor the HARPS-TERRA measurement show a
significant offset on that particular epoch.
Instead, the largest CCF offset (7.5 m s$^{-1}$)
occurred 3 days later with no apparent counter
part in the activity indices.
For comparison, the HARPS-TERRA measurement on this
same is only 3.2
m s$^{-1}$ away the mean RV. When excluding the
RV outlier, the RMS of the CCF goes from 2.37 to
1.98 \ms and TERRA goes from 2.05 to 1.88 \ms.
The extra noise to be added in quadrature to the
TERRA RMS to match the RMS of the CCF is 1.2 \ms
considering all the measurements, or 0.6 \ms if
the outlying event is excluded. An overview of
the data sets on Proxima Cen and the other two M
dwarfs discussed in previous sections (Barnard's
and Kapteyn's) is given in Table
\ref{tab:mdwarfs_overview}.

Concerning possible signals, the HARPS-TERRA RVs
periodogram show a very marginal peak at 5.6 days
that is also barely visible in the CCF values when
the outlying RV measurement is removed. So, unfortunately, no
promising signals are yet detected on Proxima Cen.

\subsection{GJ 676A, an M1V with an eccentric gas giant}

\begin{figure*}[tb]
\centering
\includegraphics[width=5.0in,clip]{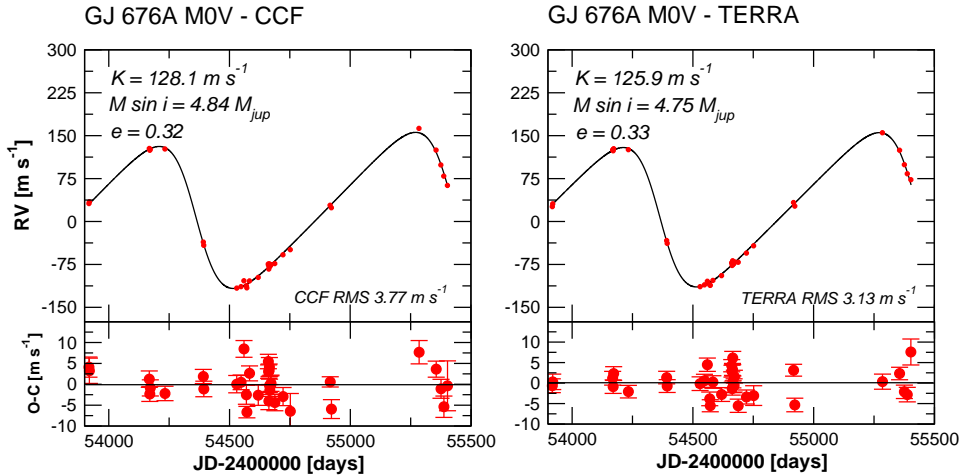}

\caption{Measured radial velocities using CCF
left and HARPS-TERRA on the right. Only the 34
epochs in the archive coincident with the
published RVs are used for this fit. This figure
clearly shows that the observed reduction in the
RMS of M dwarfs is not a spurious effect of the
data analysis techniques we apply. }

\label{fig:GJ676A_compare}
\end{figure*}

In the previous sections, we have illustrated the
improvement on the precision thanks to the use of
the template matching technique on stars with no
planets. To illustrate that the increased
precision is not an artifact of the data reduction
process, we now analyze the spectra of a planet
hosting M dwarf with a large amplitude signal. GJ
676A is orbited by a gas giant candidate with a
period of 1060 days and a RV semi-amplitude of
$\sim$ 120 m s$^{-1}$. GJ 676Ab was announced by
\citet{gj676A} and is one of the few gas giants
detected around low mass stars. The RV
measurements provided by the HARPS DRS contain two
epochs where the CCF failed to converge giving
spurious offsets of 38 km s$^{-1}$ and 76 km
s$^{-1}$ respectively. The HARPS-TERRA
measurements in these two epochs look perfectly
reasonable. For comparison purposes, these two
points were removed from the orbital analysis. As
seen in figure \ref{fig:GJ676A_compare}, the
HARPS-TERRA RVs show essentially the same RV
signal as the CCFs, demonstrating that our fitting
procedure is not knocking out real RV offsets and
that our RMS estimates on stable stars are
realistic.

After subtracting the best fit solution for one
planet, the CCF RVs shows an RMS of 6.31 \ms,
while the HARPS-Terra ones have an RMS of 6.08
\ms. Given that the typical photon noise is of the
order of 1-2 \ms, it is obvious that something
else is happening on this star. As suggested by
\citet{gj676A}, we added a linear trend to the fit
and adjusted all the free parameters again. The
improvement on both fits is quite significant
(HARPS-Terra RMS is 3.13 \ms and the RMS from the
CCF values is 3.77 \ms, see Table
\ref{tab:gj676A_fit}). As already noted by
\citep{gj676A}, the fact that the detected trend
is significantly larger than the maximal expected
acceleration due to GJ 676B strongly suggests the
presence of an additional companion to GJ 676A
with a period of several thousand days at least.
Table \ref{tab:gj676A_fit} contains the best fit
to the CCF and TERRA data sets. Only those epochs
(34 of them) that are present in both data sets
are included in both fits. The uncertainties
correspond to the 68\% confidence intervals as
obtained using the Monte Carlo Markov Chain
algorithm included in SYSTEMIC. The MCMC
jump lengths $\beta$ of each parameter were tuned
so the acceptance rate of all parameters was
between 15 an 30\% \citep{ford:2005} and 5 10$^6$
MCMC iterations were used to generate the desired
distributions and confidence levels in Table
\ref{tab:gj676A_fit}. The same data analysis
technique will be used to characterize the
uncertainties in the orbital parameters given in
the forthcoming sections.

\begin{deluxetable}{lrr}  
\tablecolumns{3}

\tablecaption{GJ 676A orbital solution using CCF
and TERRA RVs. The numbers in parenthesis
indicate the uncertainty in the last two
significant digits of the parameter values.
Uncertainties have been obtained using a Monte
Carlo Markov Chain approach (see text). }

\tablehead{
  \colhead{Parameter} &
  \colhead{CCF} &
  \colhead{TERRA}
}
\startdata
P [days]                            & 1061.7 (2.3)   & 1060.2 (1.8) \\
K [m $s^{-1}$]                      & 128.10 (66)    & 125.97 (58)  \\
M$_0$ [deg]                         & 210.13 (89)    & 208.77 (72)  \\
e                                   &  0.326  (92)   & 0.331  (78)  \\
$\omega$[deg]                       &  88.8 (1.5)    & 89.8   (1.2) \\
Linear trend [\ms yr$^{-1}$]        &  8.48 (66)     & 8.99   (56)  \\
$M \sin i$ [$M_{jup}$]              &  4.842 (25)    & 4.752  (22)  \\
a [AU]                              &  1.81          & 1.81         \\
\\
\hline\\
RMS [\ms]                           &  3.77          &   3.11       \\
$\sigma_{O-C}$[\ms]\tablenotemark{a}&  3.51          &   2.97       \\
Extra noise \tablenotemark{b}       & +2.13          &   -          \\
$N_{\rm obs}$                       &  34            &  34          \\
$N_{\rm par}$                       &   8            &   8          \\
$\chi^2$                            & 113.95         & 114.38       \\
$\chi^2$/($N_{\rm obs}-N_{\rm par}$)&   4.38         &   4.39       \\
\enddata
\tablenotetext{a}{Weighted RMS of the residuals as computed by \citet{pepe:2011}}
\tablenotetext{b}{Uncertainty that has to be added in quadrature to the CCF
RMS to match the RMS of the TERRA measurements.}

\label{tab:gj676A_fit}
\end{deluxetable}

As we have seen on the other \textit{stable} M
stars, the post-fit RMS from TERRA is
significantly lower than the one obtained with
CCF, giving further proof that, for M dwarfs,
template matching significantly outperforms CCF.
In this case, the uncertainty that has to be added
in quadrature to the HARPS-TERRA measurements is
2.1 \ms, which is very significant. As noted in
the discovery paper of GJ 676A, further low mass
planets could be present around GJ 676A in shorter
periods. Given the significantly better accuracy,
it is likely that such signals could already be
detected applying HARPS-TERRA to the complete data
set. However, even though these results were already
published, only 38 spectra are available through the
data base compared to the 69 used in \citet{gj676A},
so the full reanalysis cannot be done here. Also, we
found that several RV measurements are missing in
\citet{gj676A}. We checked that these missing
measurements coincide most of the time with the
outliers found in the archive. By comparison, no
spectrum failed to provide a useful RV measurement
using HARPS-TERRA demonstrating the robustness of
the algorithms.

\subsection{$\epsilon$ Eridani, active K3V with a gas giant
companion?}\label{sec:epseri}

$\epsilon$ Eridani is a K3V star of
approximately 0.82 M$_\sun$
\citep{butler:2006} and a close neighbor to
the Sun (3.2 pc). $\epsilon$ Eridani was first
reported to be RV variable by
\citep{campbell:1988} and \citep{walker:1995}
explicitly reported a possible variability
with a period between 5 and 10 years. A 6.9
years periodicity was first reported in
\citet{cumming:1999} but, at that time, it was
considered suspicious given the high
chromospheric activity on the star. Using
additional RV measurements,
\citet{hatzes:2000} proposed that the observed
6.9 years variability was most likely caused
by the presence of a planet. The proposed
planet would have a significantly eccentric
orbit (e$\sim$0.7) an had a minimum mass of
0.86 $M_{jup}$. The reanalysis of several RV
data sets combined with Fine Guidance
Sensor/Hubble Space Telescope astrometry
indicated that that the companion was an
actual planet with a true mass of $\sim$ 1.5
$M_{jup}$\citep{benedict:2006}. We want to
remark that the RMS of those early RV
measurements was around 10--20 \ms while the
amplitude of the claimed candidate is about 18
\ms. Also, the astrometric amplitude reported
in \citep{benedict:2006} is very close the
epoch-to-epoch systematic errors of FGS/HST.
Because of the bias towards large amplitudes
affecting astrometric measurements
\citep{pourbaix:2001}, such astrometric measurement
should be understood as an upper limit.
Evidence of gaps in the debris around
$\epsilon$ Eridani suggest an additional very
long period companion (not yet confirmed)
which should have a negligible RV signature
\citet[e.g.][]{backman:2009}. $\epsilon$
Eridani is significantly more active than the
Sun and has a strong stellar magnetic field
\citep{epseri:magnetic}. Magnetic related
activity (flaring, bright and dark spots,
etc), is the supposed source of the observed
excess in RV variability in time-scales of weeks.
Stellar global magnetic cycles could also be the
responsible for part of the observed long term
variability.


The star was regularly observed by HARPS
between 2004 and 2007. The archive contains
113 spectra of $\epsilon$ Eri. Ten spectra
showed anomalous shapes giving poor mismatches
at the order of 20\% in several echelle
apertures. A closer look did show severe
extraction problems on all of them. Three of
them are from Nov 6, 2003; and three more are
from Feb 4, 2004, all showing an anomalous
flux deficits between 5500 and 6000 \AA. The
other four are from Sep 13, 2004 and Nov 02,
2004 and all show very bumpy blaze function
shapes probably related to extraction issues
(according to the headers, these spectra were
obtained on engineering time). After excluding
these 10 measurements, the final sample we use
contains 103 HARPS observations on 23
different nights.

\begin{figure}[tb]
\centering
\includegraphics[width=2.0in,clip]{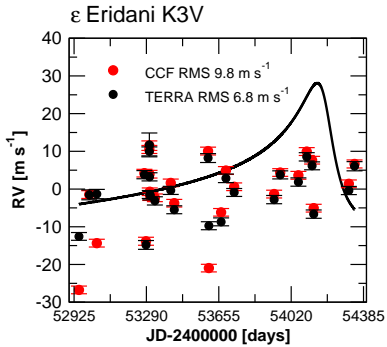}

\caption{23 nightly averages of the 103
spectra used to measure the RVs. Red symbols
represent the CCF measurements and the
HARPS-TERRA ones are shown as solid black
dots. No coherent RV variability coincident
with previous orbital solutions is observed in
either data set.} \label{fig:EpsEri_pre}

\end{figure}

The CCF measurements as extracted from the
archive have an RMS of 9.1 \ms, while the
photon noise of each observation is of the
order of $\sim$1 \ms. Such variability shows
no temporal coherence (no peaks in the
periodogram) and therefore these observations
seem to confirm that the stellar activity is
contributing significantly to the observed
scatter in the RVs. The spectra were processed
using the standard HARPS-TERRA setup for K
dwarfs. The resulting RV measurements show a
remarkably lower RMS (6.8 \ms) than the CCF
RVs. In Figure \ref{fig:EpsEri_pre}, we
present the nightly averages (23 equivalent
epochs) overlapped to the nominal solution of
planet b as given by \citet{benedict:2006}. In
the case of the CCF, when the proposed planet
is subtracted, the RMS goes from 9.3 to 11.7
\rm and the HARPS-Terra RMS increases from 6.8
to 10.0 \ms. Based on this it appears that the planet
is not real, or a major revision of its
orbital solution needs to be obtained.

\begin{figure}[tb]
\centering
\includegraphics[width=5.0in,clip]{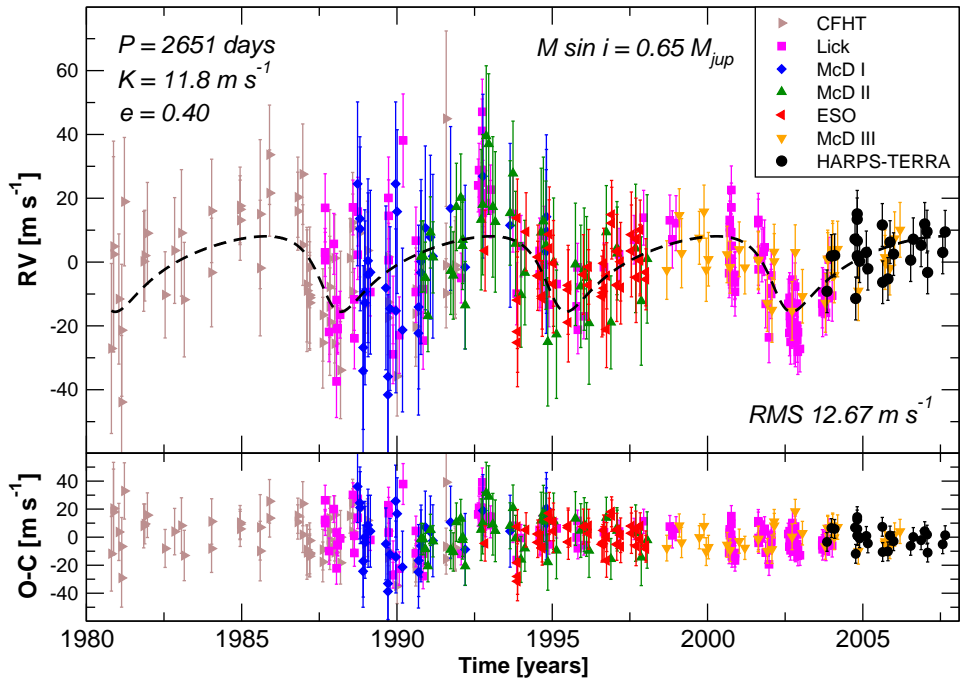}

\caption{Best orbital solution satisfying all
the radial velocities publicly available to
the date on $\epsilon$ Eridani. This solution
is significantly different to the one
presented in previous works. The new
HARPS-TERRA measurements are the black dots on
the right. The naming of the data sets follows
the prescriptions given in
\citet{benedict:2006}.}
\label{fig:EpsEri_all}

\end{figure}

Because of its proximity to the Sun and its
brightness, $\epsilon$ Eridani has been
observed by several programs with different
instruments over the years. The six data sets
available to the date are provided in the
current distribution of SYSTEMIC and were
extracted from \citet{benedict:2006} and
\citet{butler:2006}. Let us note that, because
the data used in \citep{butler:2006} was
restricted to fewer measurements, the orbital
solution presented there (only RV) was already
quite different from the one reported by
\citet{hatzes:2000} and \citet{benedict:2006}.
In order to check if there is still an orbital
solution compatible with the new RV data, we add
the new HARPS-TERRA measurements to a
Keplerian fit with the other 6 data sets. In
addition to the 5 Keplerian parameters
(period, minimum mass, eccentricity, initial
mean anomaly and argument of the node), the
model must include 7 constants to account for
the zero--point of each instrument. We make a
first tentative fit using the nominal reported
uncertainties. The obtained solution is
significantly different to the one reported by
\citet{hatzes:2000}, \citet{benedict:2006} or
\citet{butler:2006}. In particular, the new
radial velocity semi-amplitude is only $\sim$
11 \ms compared to the $\sim$18.0 \ms given on
all the previous studies and the orbital
period changes from 6.9 to 7.25 yr. This
changes are driven by the lack coherent
variability observed in the new HARPS-TERRA RVs.
The obtained eccentricity is slightly smaller
and the argument of the node is also forced to
very different values (see table
\ref{tab:epseri}). Especially in the most
recent data-sets with smaller formal
uncertainties, it is obvious that the RV
scatter is dominated by stellar noise
rather than photon noise. Since the
HARPS-TERRA measurements show the smaller RMS,
we use them to estimate the amount of stellar
jitter that has to be added in quadrature to
the nominal uncertainties in order to recover
the obtained scatter. We find that this jitter
amounts to $\sim 6.6$ \ms. By adding this 6.6
\ms in quadrature to all the reported
uncertainties, we derive a more realistic
orbital solution and corresponding
uncertainties. The definitive orbital
fit is quite similar to the one obtained
without adding the jitter but has a $\chi^2$
per degree of freedom of 1.15, indicating that
$6.6$ \ms is a reasonable estimate for the
stellar jitter. Figure
\ref{fig:EpsEri_all} and Table
\ref{tab:epseri} present the best fit to the 7
RV data-sets with the jitter included in the error bars. Even if
this orbit seems a good fit to all the
observations, we need to remark that the
orbital solution is significantly different
than the previously reported ones (e.g. the
period $P$ is at 4.2$-\sigma$ from previous
estimates). We find this difference very
suspicious and seem to indicate that the long
term RV variability of $\epsilon Eri$ is due
to stellar activity cycles (non-strictly
periodic) rather a putative planet. Let us
note that, even if the planet is real, the
astrometric measurement of the orbital
inclination and mass reported by
\citet{benedict:2006} are no longer valid due
the much more weight of the RV measurements in
the determination of the orbit. In the light
of this result, and given that significant
efforts are being devoted to attempt direct
imaging of this candidate
\citep[e.g.][]{epseri:imaging,
epseri:imaging2}, a reassessment of the
allowed orbital parameters combining
astrometry with RV measurements using modern
Monte Carlo techniques
\citep[e.g.][]{reffert:2011,anglada:2011} is
mostly needed. Also, a few additional HARPS
observations fully covering the putative
period should be sufficient to confirm that,
at least, the observed variability is still
present over a full orbital period. Precision
radial velocities in the near infrared over a full
orbital period should provide definitive
confirmation/refutation of the existence of
such planet.

\begin{deluxetable}{lr}  
\tablecolumns{3}

\tablecaption{Orbital solution for $\epsilon$
Eridani including 7 data sets. A stellar jitter
of 6.6 \ms was added in quadrature to all nominal
uncertainties. Parameter values represent the
least-squares solution to the last two
significant digits. The numbers in the
parenthesis represent uncertainty in the last two
significant digits. The uncertainties represent
the 68\% confidence level intervals as obtained
from a Bayesian MCMC.}

\tablehead{
  \colhead{Parameter} &
  \colhead{CCF}
}

\startdata
P [days]                            & 2651      (36)    \\
K [m $s^{-1}$]                      &   11.8    (1.1)    \\
M$_0$ [deg]                         &   09      (12)    \\
e\tablenotemark{a}                  &    0.40   (11)    \\
$\omega$[deg]                       &  141.4    (9.8)   \\
$M \sin i$ [$M_{jup}$]              &    0.645  (58)    \\
a [AU]                              &  3.51             \\
\\
\hline\\
RMS [\ms]                           &  12.6          \\
$\sigma_{O-C}$[\ms]\tablenotemark{b}&  10.7          \\
$N_{\rm obs}$                       &  359            \\
$N_{\rm par}$                       &  12            \\
$\chi^2$                            & 401.02         \\
$\chi^2$/($N_{\rm obs}-N_{\rm par}$)&   1.15         \\
\enddata
\tablenotetext{a}{Asymmetric distribution. $0.20< e <0.68$ with a 99\% confidence level}
\tablenotetext{b}{Weighted RMS of the residuals as computed by \citet{pepe:2011}}

\label{tab:epseri}
\end{deluxetable}

\begin{figure}[tb]
\centering
\includegraphics[width=2.0in,clip]{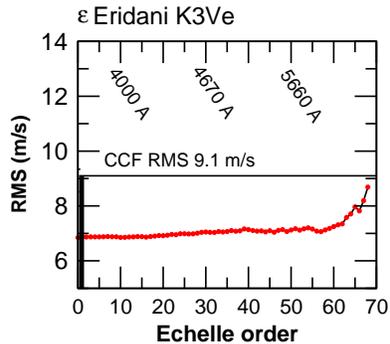}

\caption{RMS vs bluest aperture used in the
RV measurement. Even though it is an active
star, the RMS always decreases as bluer
apertures are included indicating that the
the chromatic component of the RMS is not a
dominant source of RV jitter. The CCF
measurements also use all the apertures.}

\label{fig:EpsEri_rms}
\end{figure}

In overview, the precision in the RV obtained
using the TERRA code provides a significant
reduction on the measured RMS compared to the
public CCF RVs. Given that $\epsilon$ Eridani
is an active star, the stellar lines used in
the CCF might be peculiar when compared to
quiet K dwarfs, contributing to the excess of
scatter. Even though stellar activity is the
most likely cause for most of the observed
jitter, we do not see the chromatic jitter
effect seen on M dwarfs or Tau Ceti. Figure
\ref{fig:EpsEri_rms} shows how the RMS always
decreases as bluer apertures are included.
This indicates that the leading effect
causing the observed RV jitter on active
stars is quite different to the one affecting more
quiet dwarfs.

\subsection{HD 69830, three Neptunes and
a long period trend?} \label{sec:HD69830}

HD 69830 made it to the news in 2006
\citep{lovis:2006} being the first planetary
system hosting 3 Neptune mass planets. HD
69830 is a G8V star slightly less luminous and
less massive than the sun (0.6 $L_\sun$ and
0.86 $M_\sun$ respectively). The detection was
based on 74 HARPS epochs taken between October
2003 and January 2006. The final fit to the
solution had an RMS below 1.0 \ms.

\begin{figure}[tb]
\centering
\includegraphics[width=5.0in,clip]{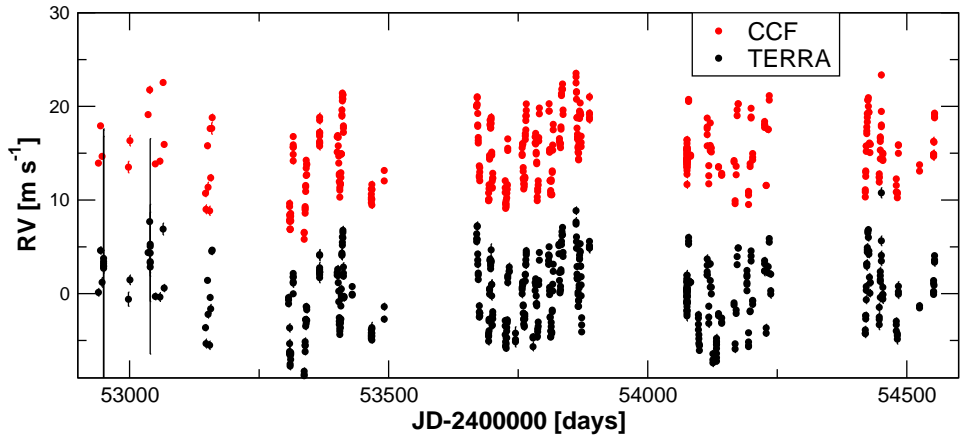}

\caption{527 Radial velocity measurements
obtained with HARPS-TERRA (black) compared to the
454 RVs obtained using a K5 binary mask with the
CCF method (red circles). The CCF RVs are shifted
15 \ms\ to improve visualization.
}\label{fig:HD69830_K5} \end{figure}

The ESO archive contains 529 spectra of the star.
The headers of two spectra were corrupted
rendering those two observations unusable (both
observations are from Feb 5th, 2007). The
resulting 527 spectra were taken between October
2003 and March 2008. The SNR at 6000\AA\ ranges
from 100 to 300 and the exposure time varies
between 180 to 400 seconds. The spectra were
processed using the standard HARPS-TERRA setup
for G dwarfs. By comparison, the CCF RV show
several outliers with RV offsets of several k\ms
away from the average(9 spectra on 3 different
nights). When removed, the RMS of the full CCF
sample is still rather high compared to the TERRA
one ($\sim$ 6.74 \ms). By direct inspection of
the CCF RV measurements, we could see that several
RVs show negative offsets of the order of 20 \ms. We found that all such
measurements (67 of them) correspond to
spectra processed by the HARPS DRS with a G2
binary mask instead of the K5 binary
mask used for the majority of the observations
(454 of them). Because the HARPS DRS is not public,
we could not consistently reobtain all the CCF
RVs. We show all the RV measurements obtained
with HARPS-TERRA and the CCF ones obtained using
the K5 binary mask in Figure
\ref{fig:HD69830_K5}.

The purpose of this section is to illustrate that
HARPS-TERRA can detect signals at the limit of the
demonstrated HARPS stability. Given the general good
agreement among the two data-sets and for the sake
of simplicity, only the HARPS-TERRA measurements
will be discussed on all that follows.

In order to perform the orbital analysis, we
consolidate these 527 observations in 176 nightly
averages. The periodogram of these RVs
directly shows the three signals reported by
\citep{lovis:2006}. The best fit solution to
these three planets compares well to the
solution given in the discovery paper and has
an RMS of 1.1 \ms.

\begin{figure}[tb]
\centering
\includegraphics[width=4.0in,clip]{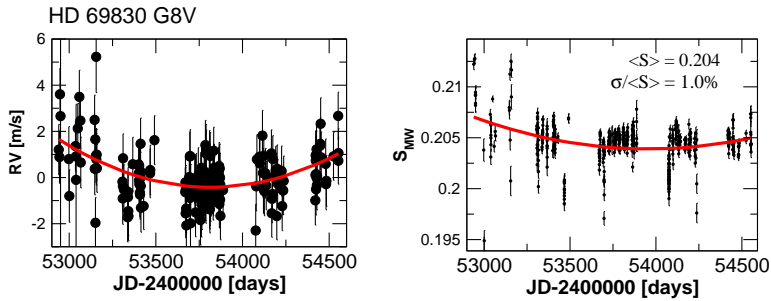}

\caption{\textbf{Left.} RV residuals to a
fully Keplerian fit to the planets already
reported by \citet{lovis:2006}. A significant
quadratic trend (red line) is observed in the
data. As reported by \citep{lovis:2006}, the
early measurements (left) have larger
uncertainties due to lower SNR and other
instrumental issues. \textbf{Right.} S index
measurements as obtained from the original 529
spectra. A clear quadratic trend is also
observed on this index pointing towards
activity as the origin of the observed long
term variability in the
RV.}\label{fig:HD69830_3postfit} \end{figure}

\begin{figure*}[tb]
\centering
\includegraphics[width=4.0in,clip]{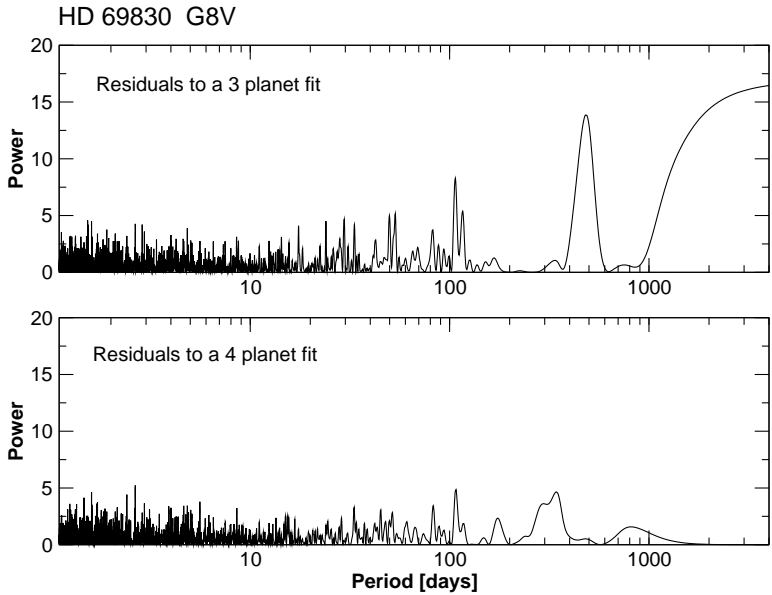}

\caption{Periodogram to the residuals of HD
69830 to a 3 planet fit (top) and to a 4
planet fit(bottom). Even though there is a
suggestive periodicity around 490 days, this
signal is most likely an alias of the long
period parabolic trend seen in  Figure
\ref{fig:HD69830_3postfit}. When the best
circular orbit to the long period trend is
removed, no additional signals are present in
the data. }\label{fig:HD69830_resper}

\end{figure*}

However, the periodogram to the residuals to
the 3 planet fit still shows a strong peak at
490 days (see Figure \ref{fig:HD69830_resper})
and an even higher power beyond 1000 days. By
visual inspection of the residuals (see Figure
\ref{fig:HD69830_3postfit}, one can see that
the long period power is due an apparent
\textit{parabolic} shape in the RV residuals.
When we tried solving for the 490 days signal,
we found that the coverage in phase of such
solution was very spotty. This is
characteristic of long period signals aliased
with the seasonal availability of the star
\citep[see][for a detailed discussion of the
yearly alias]{dawson:2010}. Given that the
power at longer periods is higher than the 490
days peak, from now on we assume that the long
period trend is the most likely signal left in
the data.

To assess what are the chances of this signal
being generated by random noise, we compute
its empirical false alarm probability as
follows. We generate 10$^4$ synthetic data
sets by keeping the observed epochs but doing
random permutations of the residuals to the 3
planet fit. While these synthetic data sets
will have the same distribution of random
errors, the random permutations destroy the
temporal coherence of any signal present. We
then compute the periodogram for each
synthetic data set and count the number of
times we obtain a power higher than the one we
find in the real data. This only happened once
indicating that the FAP is low ($\sim$0.01 \%),
which means that such trend is statistically
significant and cannot be ignored.  This method
of computing FAP is described in more detail
in \citet{cumming:2004}.

In order to assess what could be the cause of
this trend, we examined the measurements on
the S-index as described in Section
\ref{sec:svalue}. Unfortunately, the S-index
shows a trend with the same shape and relative
variability as the RV curve. This is
characteristic of spurious offsets caused
by the magnetic activity cycle of the star
and has been observed in other stable stars
such as Tau Ceti \citep{pepe:2011}. A
tentative fit of the RV residuals provides a
period of  $\sim 8500\pm 2000$ which roughly
matches the expected duration of the activity
cycle of a quiet star similar to our Sun. Even
though the signal could be the combination of
a long period planet and the activity induced
offset, we cannot conclude that there is solid
evidence for an additional companion to HD 69830.
Once a circular orbit is subtracted to
the data, the signal at 490 days completely
disappears from the periodogram, confirming
that both peaks in Figure
\ref{fig:HD69830_resper} correspond to the
same aliased trend.

\begin{figure*}[tb]
\centering
\includegraphics[width=3.0in,clip]{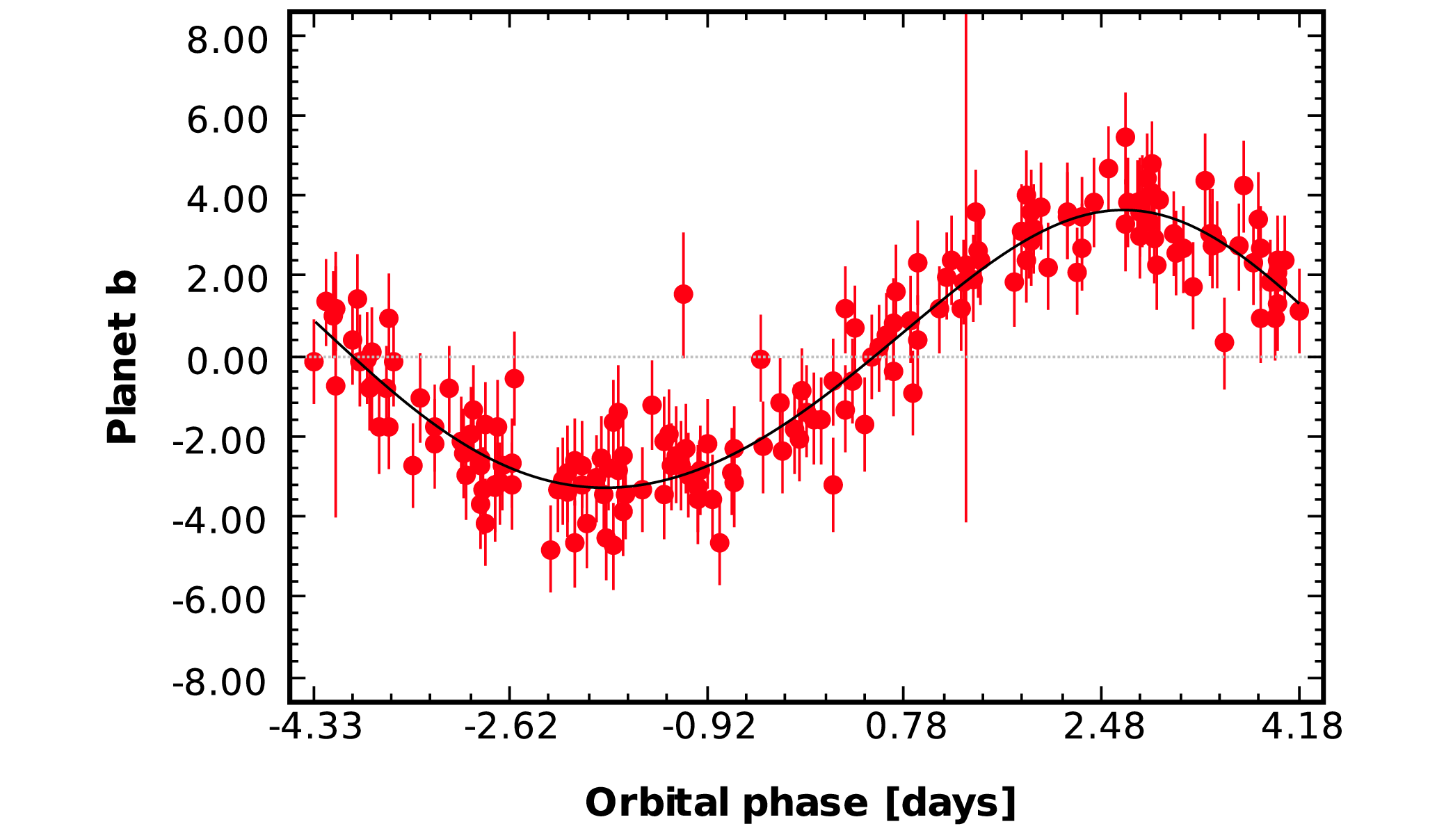}
\includegraphics[width=3.0in,clip]{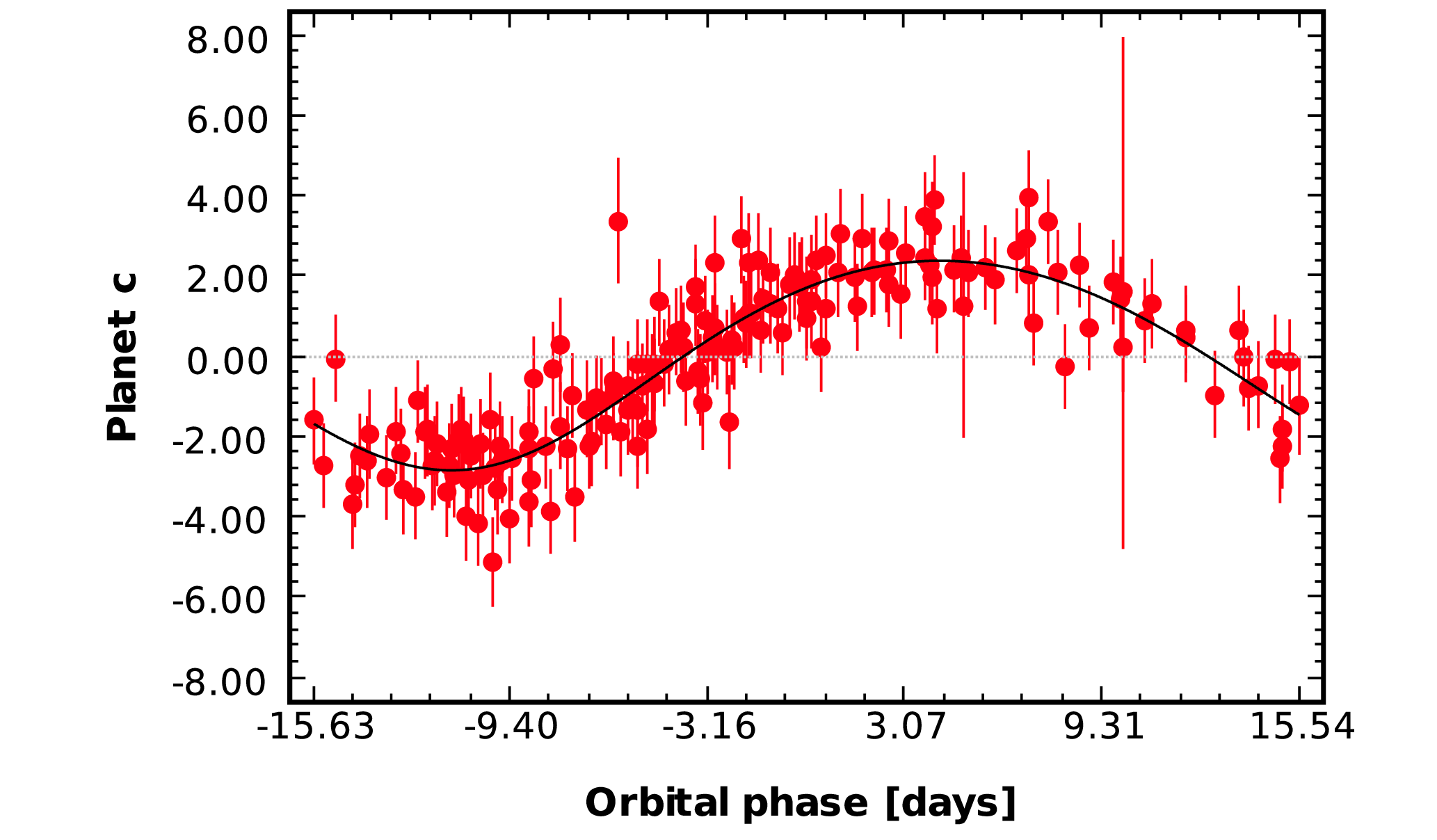}
\includegraphics[width=3.0in,clip]{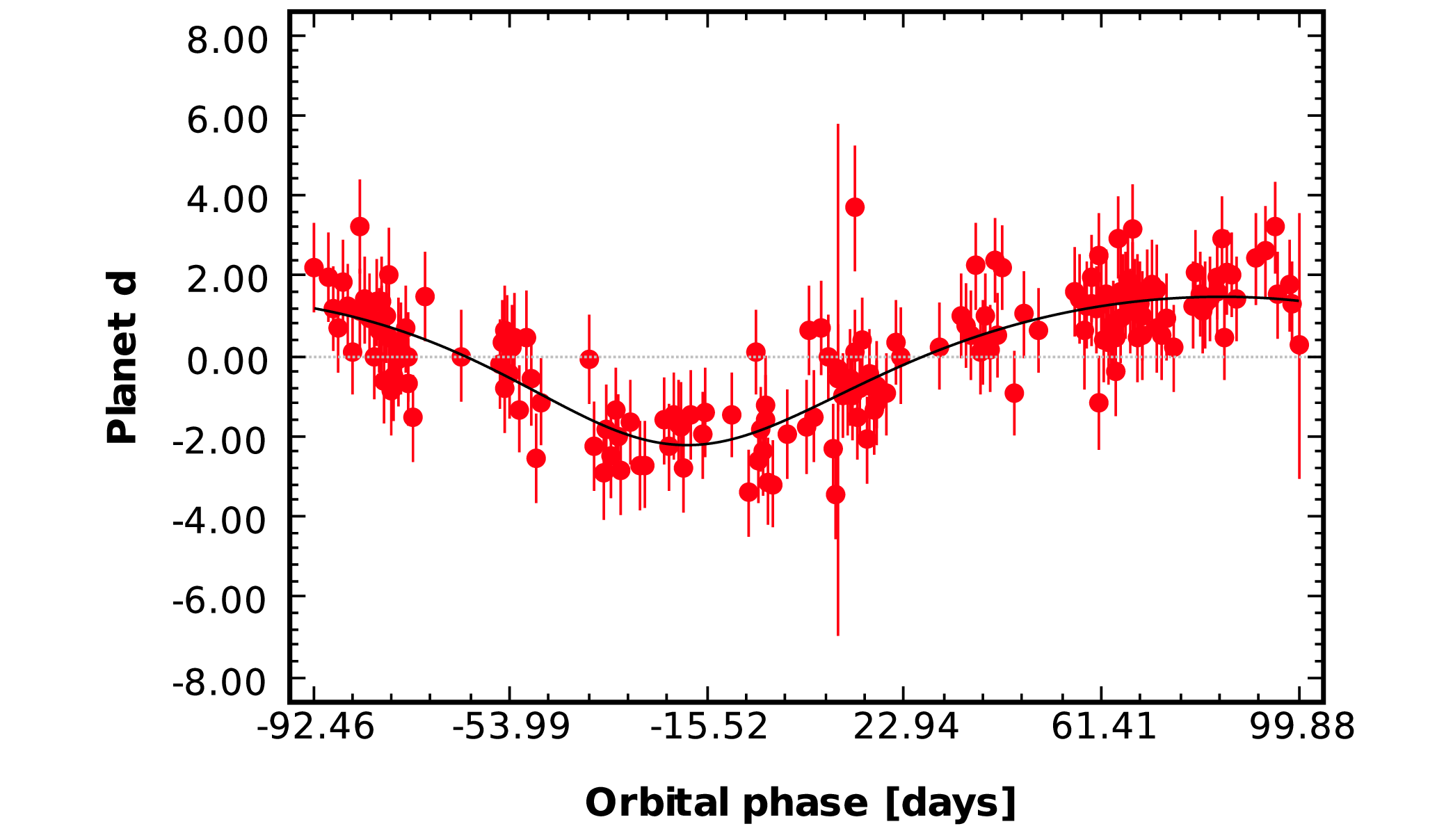}
\includegraphics[width=3.0in,clip]{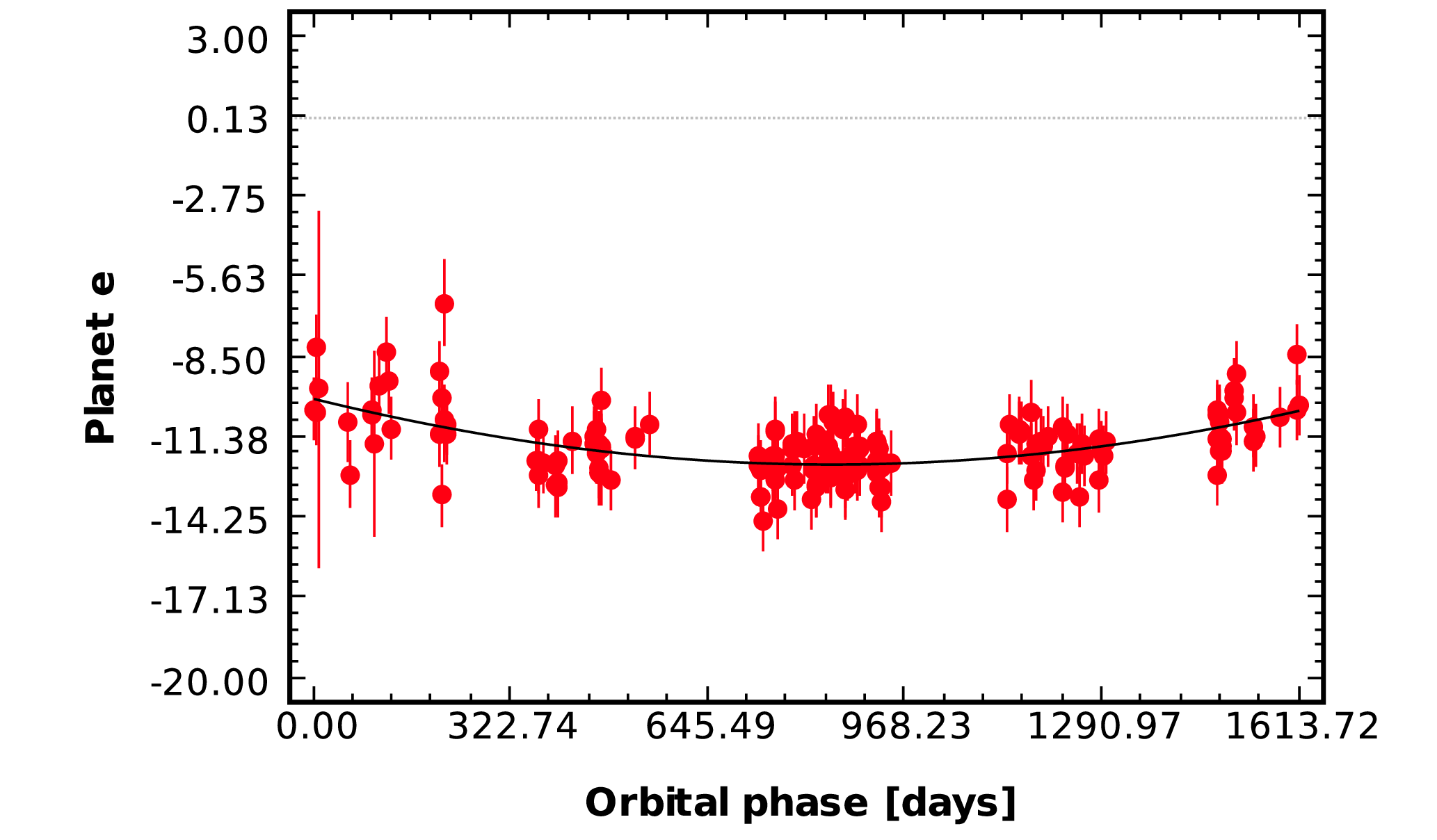}

\caption{Best fit to the 4 signals present in the HD 69830
measurements obtained with HARPS-TERRA. The first three
planets were already reported by \citep{lovis:2006}. A
circular orbit for planet e has been assumed to generate this
plot. The RMS of the residuals to the 4 signals fit is only
0.92 \ms }

\label{fig:HD69830_fit}
\end{figure*}

\begin{deluxetable}{lrrrr}  
\tablecolumns{5}

\tablecaption{Orbital solution for the 3 planet
candidates plus trend detected on the RVs of HD 69830 when
using all the echelle apertures.
The proposed parameter
values are obtained from the $\chi^2$ solution.
The numbers in parenthesis correspond to the uncertainty
in the last two significant digits in the parameter values
(68\% confidence level intervals). Statistical quantities at
the bottom correspond to a circular orbital fit to
the quadratic trend. All the orbital elements are
referred to the first epoch of observation at
JD$_0$=2452939.87402 days.}

\tablehead{
  \colhead{} &
  \colhead{b} &
  \colhead{c} &
  \colhead{d} &
  \colhead{(activity?)}
}
\startdata
P [days]                          & 8.6687 (12)   & 31.645 (28)   & 202.2 (1.6)    & 8500\tablenotemark{a}\\
K [\ms]                           & 3.46 (11)     & 2.61 (22)     & 1.85  (13)	   & 12.35\tablenotemark{a}         \\
$M_0$ [deg]                       & 228 (10)      & 118 (15)      & 39    (17)     & 144\tablenotemark{a}           \\
e                                 & 0.06$^*$ (05) & 0.08$^*$ (05) & 0.190$^+$ (76) & 0 (fixed)     \\
$\omega$[deg]                     & 24 (23)       & 192 (40)      & 172   (32)     & 0 (fixed)	 \\
$M \sin i$ [$M_{jup}$]            & 0.0315 (11)   & 0.0366 (22)   & 0.0473 (35)    & 1.12\tablenotemark{a}	    \\
$M \sin i$ [$M_{\Earth}$]         & 10.00(32)     & 11.60 (69)    & 15.0  (1.1)	   & 355.4\tablenotemark{a}	    \\
a [AU]                            & 0.079         & 0.186         & 0.641          & 7.7 AU\tablenotemark{a}      \\
\\
\hline\\
RMS [\ms]                              &   0.92     \\
$\sigma_{O-C}$[\ms]\tablenotemark{b}   &   0.88     \\
$N_{\rm obs}$                          &    176     \\
$N_{\rm par}$                          &     19     \\
$\chi^2$                               & 120.35     \\
$\frac{\chi^2}{N_{\rm obs}-N_{\rm par}}$   &   0.76     \\
\enddata

\tablenotetext{a}{Orbital values of the equivalent circular
orbit. Note that the most likely explanation to this signal is
the RV offset induced by the magnetic cycle of the star.}

\tablenotetext{b}{Weighted RMS of the residuals as computed by \citet{pepe:2011}}
\tablenotetext{*}{Compatible with circular orbit}
\tablenotetext{+}{Slightly significant eccentricity}
\label{tab:HD69830_solution}
\end{deluxetable}


The number of observations used here is
significantly larger than those available at the
time of discovery of the first 3 candidates.
Also, only the long time-span of the
observations ($\sim$ 5 years) allows to detect
the quadratic trend and its correlation with the
S-index variability. As shown in the bottom
panel of Figure \ref{fig:HD69830_resper}, no
more periodicities can be inferred from the
residuals to the 4 signal solution. Removing the
trend with a circular orbit leaves an RMS 0.92
\ms. Note that, even with the trend, this is one
of the most RV stable G dwarfs observed by
HARPS. As a final note, we analyzed the CCF RV
obtained with the K5 binary mask using the same
procedure. We recovered the same orbital
solution for the three reported candidates, but
the long period trend and its alias at 490 days
appear both with less significance. While one
would be tempted to claim that HARPS-TERRA
obtains higher precision also in this case, this
extreme cannot be confirmed here for three
reasons : 1) the CCF data set contains fewer
measurements, 2) the RMS of both measurements
after removing the first 3 planets is almost
identical, 3) such precision is at the limit of
the long term stability of HARPS and any signal
at this level (specially long period ones) has to
be taken with due caution.

\subsubsection{Wavelength dependence of the
signals in HD 69830}\label{sec:HD69830_wave}

\begin{figure*}[tb] \centering
\includegraphics[width=3in,clip]{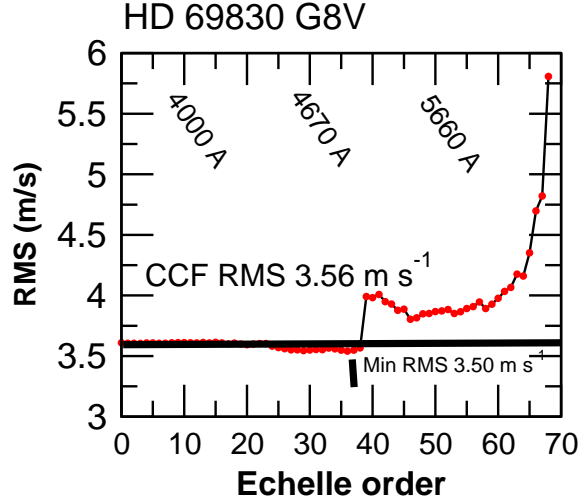}

\caption{Radial velocity RMS as a function of
the bluer aperture used on HD 69830. Even
though this raw RMS contains the signal of three
low amplitude companions, the contribution of
the Keplerian signals and the chromatic jitter
should add in quadrature, providing an
optimal bluer aperture to be used.}
\label{fig:MinOrder_HD69830}
\end{figure*}

We performed an additional test to assess the
reality of the aforementioned trend. Given the
similar spectral type of HD 69830 (G8V) to Tau
Ceti (G8.5V) we asked ourselves if the chromatic
jitter effect detected in the RMS of M dwarfs
and Tau Ceti could be exploited to investigate
the nature of the trend. To do this, we first
plotted the RMS as a function of the bluer
aperture used and found its minimum (see Figure
\ref{fig:MinOrder_HD69830}). Let us note that,
although there are three low amplitude
candidates contributing to the initial RMS,
Keplerian signals should be achromatic
and, therefore, a minimum in the RMS should
still be present when the wavelength dependent
noise is added in quadrature. Figure
\ref{fig:MinOrder_HD69830} shows that such
minimum in the RMS is found at aperture 37,
which is similar to the optimal bluer aperture
found for Tau Ceti.

\begin{figure*}[tb]
\centering
\includegraphics[width=4.0in,clip]{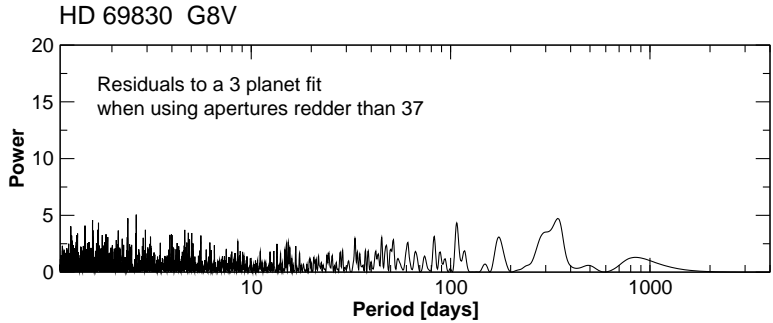}

\caption{Periodogram to the residuals of HD
69830 to a 3 planet fit (top) when only
apertures redder than 37 are used to produce the
RV measurements. No hint of the trend or any other
periodicity is seen in the data.}
\label{fig:HD69830_O37_resper}
\end{figure*}

We then used the RVs obtained using only apertures
redder than 37 and derived the full Keplerian
solution for the first three planets candidates.
The obtained orbits were compatible with the previously
reported orbits. However, when we computed the periodogram of
the residuals to the three planet fit, we found that
the quadratic trend and the corresponding 490 days
alias were completely gone. Also, the RMS to the 3
planet fit was already very low (1.01 \ms), clearly
indicating that the secular signal was mainly driven
by apparent RV offsets at the bluest wavelengths.
Even though instrumental effects might be involved,
we think this result strongly suggests that, 1) this
chromatic jitter is intrinsic to the star, 2) it is a
significant source of the stellar RV noise observed
in other relatively quiet stars, and 3) might pose a
fundamental limit to the maximum RV precision
achievable on G stars.

\section{Summary and conclusions} \label{sec:conclusions}

We present a new method to obtain precision radial
velocities from wavelength calibrated public HARPS
spectra. Our new velocities compare well with those
obtained with the CCF method and are able to detect
the same signals reported by previous HARPS
discoveries. For stable G and K dwarfs, the
difference between CCF and HARPS-TERRA is at the
level of $\sim$ 0.5 \ms\, RMS. Still, the template
matching approach seems to be less sensitive to
offsets induced by the stellar activity (e.g. see
Section \ref{sec:epseri} on $\epsilon$ Eridani) and
requires almost no assumptions on the nature of the
star. Because the template matching technique makes a
more optimal use of the Doppler information in the
stellar spectrum, HARPS-TERRA provides a significant
increase in accuracy when applied to the heavily
blended spectra of M dwarfs.

We have shown the importance of correcting for blaze
function variability (flux normalization polynomial)
to achieve sub \ms precision. While the CCF method
needs some preprocessing of the spectra to account
for such variability, least-squares template matching
performs such correction in a self-consistent way.

We find that several stars show excess variability
when the bluer echelle apertures are used. Even if
the SNR is typically lower in the blue, this increase
in the \textit{jitter} should not happen on perfectly
stable stars. We find significant evidence that this
\textit{chromatic jitter} is likely related to the
star itself rather than an instrumental effect and
that the wavelength dependence of the RV offsets can
be exploited to confirm or rule out suspicious candidate
signals. Further research is necessary to assess the
nature of this excess.

We also have demonstrated that HARPS-TERRA can detect
already reported signals and we have applied it to a
number of interesting stars with abundant public
data. While we can reproduce well other
detections, the new HARPS-TERRA measurements do not
firmly confirm the planet candidate reported around
$\epsilon$ Eridani. The orbital
solution allowed by the new observations is
significantly different compared to the previously
reported ones, casting some doubts on the reality of
this candidate.

We also report the detection of a quadratic trend in
the residuals of HD 69830. Despite the high
significance of the signal, we find that it
correlates well with a similar trend in the S-index.
We also find that such trend completely disappears
when only the redder half of the spectrum is used to
derive the RV measurements. This leads us to conclude
that the most likely explanation for the observed
trend is the activity cycle of HD 69830 inducing
wavelength dependent RV shifts that are stronger towards the blue. Although we do not report any new planet, this example demonstrates that
HARPS-TERRA is also able to robustly detect and
diagnose unreported signals at the limit of the HARPS
instrumental stability. Given that HD 69830 is a
nearby star, a planet candidate with such a long
period would lie around 0.7$\arcsec$ from the central
star and might be imaged with the next generation of
adaptive optics systems
\citep[e.g.][]{lagrange:2010}. Even though everything
points to stellar activity as the most likely explanation
for the trend, we provide,just in case, an estimate of the
equivalent circular orbit.

We have shown that the template matching on
stabilized spectrometers requires few
assumptions compared to the elaborate binary masks
required by the CCF method. The capability of
reproducing precision RV measurements with two
different data analysis methods is also a powerful
ally to double check the significance of very low
amplitude signals. Given the significant increase in
precision achieved on low mass stars, it is likely
that HARPS-TERRA can uncover undetected low amplitude
signals in already existing data sets
\citep{anglada:2012a}.

\textbf{Acknowledgements} The research of GA
has been supported by the Carnegie Fellowship
Postdoctoral Program. We thank Mathias Zechmeister,
Steve Vogt, Alan Boss and Alycia Weinberger for useful
discussions. RPB gratefully acknowledges
support from NASA OSS Grant NNX07AR40G and
from the Carnegie Institution of Washington.


\begin{thebibliography}{49}
\expandafter\ifx\csname natexlab\endcsname\relax\def\natexlab#1{#1}\fi

\bibitem[{{Anglada-Escud{\'e}} {et~al.}(2012){Anglada-Escud{\'e}}, {Arriagada},
  {Vogt}, \& et~al.}]{anglada:2012a}
{Anglada-Escud{\'e}}, G., {Arriagada}, P., {Vogt}, S.~S., \& et~al. 2012,
  accepted to ApJL, --,

\bibitem[{{Anglada-Escud{\'e}} {et~al.}(2011){Anglada-Escud{\'e}}, {Boss},
  {Weinberger}, {Thompson}, {Butler}, {Vogt}, \& {Rivera}}]{anglada:2011}
{Anglada-Escud{\'e}}, G., {Boss}, A.~P., {Weinberger}, A.~J., {Thompson},
  I.~B., {Butler}, R.~P., {Vogt}, S.~S., \& {Rivera}, E.~J. 2011, ArXiv
  e-prints

\bibitem[{{Backman} {et~al.}(2009){Backman}, {Marengo}, {Stapelfeldt}, {Su},
  {Wilner}, {Dowell}, {Watson}, {Stansberry}, {Rieke}, {Megeath}, {Fazio}, \&
  {Werner}}]{backman:2009}
{Backman}, D., {et~al.} 2009, \apj, 690, 1522

\bibitem[{{Baranne} {et~al.}(1996){Baranne}, {Queloz}, {Mayor}, {Adrianzyk},
  {Knispel}, {Kohler}, {Lacroix}, {Meunier}, {Rimbaud}, \&
  {Vin}}]{baranne:1996}
{Baranne}, A., {et~al.} 1996, \aaps, 119, 373

\bibitem[{{Benedict} {et~al.}(2006){Benedict}, {McArthur}, {Gatewood}, {Nelan},
  {Cochran}, {Hatzes}, {Endl}, {Wittenmyer}, {Baliunas}, {Walker}, {Yang},
  {K{\"u}rster}, {Els}, \& {Paulson}}]{benedict:2006}
{Benedict}, G.~F., {et~al.} 2006, \aj, 132, 2206

\bibitem[{{Butler} {et~al.}(1996){Butler}, {Marcy}, {Williams}, {McCarthy},
  {Dosanjh}, \& {Vogt}}]{butler:1996}
{Butler}, R.~P., {Marcy}, G.~W., {Williams}, E., {McCarthy}, C., {Dosanjh}, P.,
  \& {Vogt}, S.~S. 1996, \pasp, 108, 500

\bibitem[{{Butler} {et~al.}(2006){Butler}, {Wright}, {Marcy}, {Fischer},
  {Vogt}, {Tinney}, {Jones}, {Carter}, {Johnson}, {McCarthy}, \&
  {Penny}}]{butler:2006}
{Butler}, R.~P., {et~al.} 2006, \apj, 646, 505

\bibitem[{{Campbell} {et~al.}(1988){Campbell}, {Walker}, \&
  {Yang}}]{campbell:1988}
{Campbell}, B., {Walker}, G.~A.~H., \& {Yang}, S. 1988, \apj, 331, 902

\bibitem[{{Cumming}(2004)}]{cumming:2004}
{Cumming}, A. 2004, \mnras, 354, 1165

\bibitem[{{Cumming} {et~al.}(1999){Cumming}, {Marcy}, \&
  {Butler}}]{cumming:1999}
{Cumming}, A., {Marcy}, G.~W., \& {Butler}, R.~P. 1999, \apj, 526, 890

\bibitem[{{Dawson} \& {Fabrycky}(2010)}]{dawson:2010}
{Dawson}, R.~I., \& {Fabrycky}, D.~C. 2010, \apj, 722, 937

\bibitem[{{Delfosse} {et~al.}(2000){Delfosse}, {Forveille}, {S{\'e}gransan},
  {Beuzit}, {Udry}, {Perrier}, \& {Mayor}}]{delfosse:2000}
{Delfosse}, X., {Forveille}, T., {S{\'e}gransan}, D., {Beuzit}, J.-L., {Udry},
  S., {Perrier}, C., \& {Mayor}, M. 2000, \aap, 364, 217

\bibitem[{{Endl} \& {K{\"u}rster}(2008)}]{endl:2008}
{Endl}, M., \& {K{\"u}rster}, M. 2008, \aap, 488, 1149

\bibitem[{{Figueira} {et~al.}(2010){Figueira}, {Pepe}, {Melo}, {Santos},
  {Lovis}, {Mayor}, {Queloz}, {Smette}, \& {Udry}}]{figueira:2010}
{Figueira}, P., {et~al.} 2010, \aap, 511, A55+

\bibitem[{{Ford}(2005)}]{ford:2005}
{Ford}, E.~B. 2005, \aj, 129, 1706

\bibitem[{{Forveille} {et~al.}(2011{\natexlab{a}}){Forveille}, {Bonfils}, {Lo
  Curto}, {Delfosse}, {Udry}, {Bouchy}, {Lovis}, {Mayor}, {Moutou}, {Naef},
  {Pepe}, {Perrier}, {Queloz}, \& {Santos}}]{gj676A}
{Forveille}, T., {et~al.} 2011{\natexlab{a}}, \aap, 526, A141+

\bibitem[{{Forveille} {et~al.}(2011{\natexlab{b}}){Forveille}, {Bonfils},
  {Delfosse}, {Alonso}, {Udry}, {Bouchy}, {Gillon}, {Lovis}, {Neves}, {Mayor},
  {Pepe}, {Queloz}, {Santos}, {Segransan}, {Almenara}, {Deeg}, \&
  {Rabus}}]{forveille:2011b}
---. 2011{\natexlab{b}}, ArXiv e-prints

\bibitem[{{Hatzes} {et~al.}(2000){Hatzes}, {Cochran}, {McArthur}, {Baliunas},
  {Walker}, {Campbell}, {Irwin}, {Yang}, {K{\"u}rster}, {Endl}, {Els},
  {Butler}, \& {Marcy}}]{hatzes:2000}
{Hatzes}, A.~P., {et~al.} 2000, \apjl, 544, L145

\bibitem[{{Heinze} {et~al.}(2008){Heinze}, {Hinz}, {Kenworthy}, {Miller}, \&
  {Sivanandam}}]{epseri:imaging2}
{Heinze}, A.~N., {Hinz}, P.~M., {Kenworthy}, M., {Miller}, D., \& {Sivanandam},
  S. 2008, \apj, 688, 583

\bibitem[{{Howard} {et~al.}(2010){Howard}, {Marcy}, {Johnson}, {Fischer},
  {Wright}, {Isaacson}, {Valenti}, {Anderson}, {Lin}, \& {Ida}}]{howard:2010}
{Howard}, A.~W., {et~al.} 2010, Science, 330, 653

\bibitem[{{Hu{\'e}lamo} {et~al.}(2008){Hu{\'e}lamo}, {Figueira}, {Bonfils},
  {Santos}, {Pepe}, {Gillon}, {Azevedo}, {Barman}, {Fern{\'a}ndez}, {di Folco},
  {Guenther}, {Lovis}, {Melo}, {Queloz}, \& {Udry}}]{huelamo:2008}
{Hu{\'e}lamo}, N., {et~al.} 2008, \aap, 489, L9

\bibitem[{{Janson} {et~al.}(2007){Janson}, {Brandner}, {Henning}, {Lenzen},
  {McArthur}, {Benedict}, {Reffert}, {Nielsen}, {Close}, {Biller}, {Kellner},
  {G{\"u}nther}, {Hatzes}, {Masciadri}, {Geissler}, \&
  {Hartung}}]{epseri:imaging}
{Janson}, M., {et~al.} 2007, \aj, 133, 2442

\bibitem[{{Kaltenegger} {et~al.}(2011){Kaltenegger}, {Udry}, \&
  {Pepe}}]{kaltenegger:2011}
{Kaltenegger}, L., {Udry}, S., \& {Pepe}, F. 2011, ArXiv e-prints

\bibitem[{{Lagrange} {et~al.}(2010){Lagrange}, {Bonnefoy}, {Chauvin}, {Apai},
  {Ehrenreich}, {Boccaletti}, {Gratadour}, {Rouan}, {Mouillet}, {Lacour}, \&
  {Kasper}}]{lagrange:2010}
{Lagrange}, A.-M., {et~al.} 2010, Science, 329, 57

\bibitem[{{Lovis} {et~al.}(2006){Lovis}, {Mayor}, {Pepe}, {Alibert}, {Benz},
  {Bouchy}, {Correia}, {Laskar}, {Mordasini}, {Queloz}, {Santos}, {Udry},
  {Bertaux}, \& {Sivan}}]{lovis:2006}
{Lovis}, C., {et~al.} 2006, \nat, 441, 305

\bibitem[{{Lovis} {et~al.}(2011){Lovis}, {Dumusque}, {Santos}, {Bouchy},
  {Mayor}, {Pepe}, {Queloz}, {S{\'e}gransan}, \& {Udry}}]{lovis:2011}
---. 2011, ArXiv e-prints

\bibitem[{{Marcy} \& {Butler}(1996)}]{marcy:1996}
{Marcy}, G.~W., \& {Butler}, R.~P. 1996, \apjl, 464, L147

\bibitem[{{Mayor} \& {Queloz}(1995)}]{mayor:1995}
{Mayor}, M., \& {Queloz}, D. 1995, \nat, 378, 355

\bibitem[{{Mayor} {et~al.}(2009){Mayor}, {Bonfils}, {Forveille}, {Delfosse},
  {Udry}, {Bertaux}, {Beust}, {Bouchy}, {Lovis}, {Pepe}, {Perrier}, {Queloz},
  \& {Santos}}]{mayor:2009}
{Mayor}, M., {et~al.} 2009, \aap, 507, 487

\bibitem[{{Mayor} {et~al.}(2011){Mayor}, {Marmier}, {Lovis}, {Udry},
  {S{\'e}gransan}, {Pepe}, {Benz}, {Bertaux}, {Bouchy}, {Dumusque}, {Lo Curto},
  {Mordasini}, {Queloz}, \& {Santos}}]{mayor:2011}
---. 2011, ArXiv e-prints

\bibitem[{{Meschiari} \& {Laughlin}(2010)}]{meschiari:2010}
{Meschiari}, S., \& {Laughlin}, G.~P. 2010, \apj, 718, 543

\bibitem[{{Pepe} {et~al.}(2002){Pepe}, {Mayor}, {Galland}, \&
  et~al.}]{pepe:2002}
{Pepe}, F., {Mayor}, M., {Galland}, \& et~al. 2002, \aap, 388, 632

\bibitem[{{Pepe} {et~al.}(2003){Pepe}, {Rupprecht}, {Avila}, \&
  el~al.}]{harps:construction}
{Pepe}, F., {Rupprecht}, G., {Avila}, G., \& el~al. 2003, in SPIE Conference
  Series, Vol. 4841, 1045--1056

\bibitem[{{Pepe} {et~al.}(2011){Pepe}, {Lovis}, {S{\'e}gransan}, {Benz},
  {Bouchy}, {Dumusque}, {Mayor}, {Queloz}, {Santos}, \& {Udry}}]{pepe:2011}
{Pepe}, F., {et~al.} 2011, \aap, 534, A58+

\bibitem[{{Pourbaix}(2001)}]{pourbaix:2001}
{Pourbaix}, D. 2001, \aap, 369, L22

\bibitem[{{Press} {et~al.}(1992){Press}, {Teukolsky}, {Vetterling}, \&
  {Flannery}}]{numerical}
{Press}, W.~H., {Teukolsky}, S.~A., {Vetterling}, W.~T., \& {Flannery}, B.~P.
  1992, {Numerical recipes in FORTRAN. The art of scientific computing}
  (Cambridge: University Press, |c1992, 2nd ed.)

\bibitem[{{Queloz}(1995)}]{queloz:1995}
{Queloz}, D. 1995, in IAU Symposium, Vol. 167, New Developments in Array
  Technology and Applications, ed. {A.~G.~D.~Philip, K.~Janes, \&
  A.~R.~Upgren}, 221

\bibitem[{{Queloz} {et~al.}(2001){Queloz}, {Henry}, {Sivan}, {Baliunas},
  {Beuzit}, {Donahue}, {Mayor}, {Naef}, {Perrier}, \& {Udry}}]{queloz:2001}
{Queloz}, D., {et~al.} 2001, \aap, 379, 279

\bibitem[{{Reffert} \& {Quirrenbach}(2011)}]{reffert:2011}
{Reffert}, S., \& {Quirrenbach}, A. 2011, \aap, 527, A140

\bibitem[{{Reiners} {et~al.}(2010){Reiners}, {Bean}, {Huber}, {Dreizler},
  {Seifahrt}, \& {Czesla}}]{reiners:2010}
{Reiners}, A., {Bean}, J.~L., {Huber}, K.~F., {Dreizler}, S., {Seifahrt}, A.,
  \& {Czesla}, S. 2010, \apj, 710, 432

\bibitem[{{Rivera} {et~al.}(2010){Rivera}, {Laughlin}, {Butler}, {Vogt},
  {Haghighipour}, \& {Meschiari}}]{rivera:2010}
{Rivera}, E.~J., {Laughlin}, G., {Butler}, R.~P., {Vogt}, S.~S.,
  {Haghighipour}, N., \& {Meschiari}, S. 2010, \apj, 719, 890

\bibitem[{{Rueedi} {et~al.}(1997){Rueedi}, {Solanki}, {Mathys}, \&
  {Saar}}]{epseri:magnetic}
{Rueedi}, I., {Solanki}, S.~K., {Mathys}, G., \& {Saar}, S.~H. 1997, \aap, 318,
  429

\bibitem[{{Standish}(1998)}]{standish:1998}
{Standish}, E.~M. 1998, JPL Tech Note, IOM 312.F-98-048

\bibitem[{{Stumpff}(1979)}]{stumpff:1979}
{Stumpff}, P. 1979, \aap, 78, 229

\bibitem[{{Stumpff}(1980)}]{stumpff:1980}
---. 1980, \aaps, 41, 1

\bibitem[{{van Leeuwen}(2007)}]{hipparcos:2007}
{van Leeuwen}, F. 2007, \aap, 474, 653

\bibitem[{{Walker} {et~al.}(1995){Walker}, {Walker}, {Irwin}, {Larson}, {Yang},
  \& {Richardson}}]{walker:1995}
{Walker}, G.~A.~H., {Walker}, A.~R., {Irwin}, A.~W., {Larson}, A.~M., {Yang},
  S.~L.~S., \& {Richardson}, D.~C. 1995, \icarus, 116, 359

\bibitem[{{Zechmeister} \& {K{\"u}rster}(2009)}]{zechmeister:2009b}
{Zechmeister}, M., \& {K{\"u}rster}, M. 2009, \aap, 496, 577

\bibitem[{{Zechmeister} {et~al.}(2009){Zechmeister}, {K{\"u}rster}, \&
  {Endl}}]{zechmeister:2009}
{Zechmeister}, M., {K{\"u}rster}, M., \& {Endl}, M. 2009, \aap, 505, 859

\end{thebibliography}

\appendix
\section{Radial velocity measurements} \label{sec:rvdata}

This appendix contains tables with the most relevant
time-series used through the paper. Unless stated
otherwise, the first two columns are the RVs as obtained
with HARPS-TERRA using the standard setup for each
spectral type. The corresponding CCF values are also
provided if necessary for comparison purposes. An offset
(average RV) has been subtracted to all the RVs to
improve readibility and all RV measurements are Doppler
offsets measured in the Solar System Barycentric system.
The perspective acceleration effect has also been
removed from all the presented RVs. No nightly averages
have been applied. The tables also contain the S-index
values in the Mount-Wilson system  as measured by
HARPS-TERRA. In a few cases, the RV measurements using
the redder part of the spectrum are provided instead the
S-index or the CCF measurements. Check each table
caption for further information.

\begin{deluxetable}{lccccccc}  
\tablecolumns{5}

\tablecaption{Measurements on Tau Ceti.}
\tablehead{
  \colhead{JD } &
  \colhead{RV$_{\rm TERRA}$} &
  \colhead{$\sigma_{\rm TERRA}$} &
  \colhead{RV$_{\rm CCF}$} &
  \colhead{$\sigma_{\rm CCF}$} &
  \colhead{S-index} &
  \colhead{$\sigma_{S}$}
  \\
  \colhead{[days]} &
  \colhead{[\ms]} &
  \colhead{[\ms]} &
  \colhead{[\ms]} &
  \colhead{[\ms]} &
  \colhead{\, } &
  \colhead{\, }
  }
\startdata
 2453280.550009 &    0.42615 &    0.51027 &    0.16668 &    0.30930 &    0.13932 &    0.00050 \\
 2453281.551756 &    1.58720 &    0.44181 &    1.69523 &    0.28281 &    0.14008 &    0.00045 \\
 2453282.551419 &    0.47161 &    0.52024 &    0.92778 &    0.33676 &    0.13800 &    0.00053 \\
 2453283.550154 &    1.18747 &    0.54740 &    0.46403 &    0.32926 &    0.13992 &    0.00052 \\
 2453284.793218 &   -0.16593 &    0.69264 &   -0.05192 &    0.60515 &    0.13798 &    0.00089 \\
 2453309.755272 &    0.03438 &    0.28136 &    0.62089 &    0.18451 &    0.14151 &    0.00028 \\
 2453336.661904 &   -1.67321 &    0.39949 &   -0.85605 &    0.19063 &    0.14184 &    0.00029 \\
 2453337.609246 &   -3.42329 &    0.31737 &   -3.25516 &    0.17864 &    0.14239 &    0.00024 \\
\enddata

\label{tab:rv_tauceti}
\end{deluxetable}

\begin{deluxetable}{lcccccc}  
\tablecolumns{5}

\tablecaption{Differential RV and S-index measurements
on HD 85512.}

\tablehead{
  \colhead{JD } &
  \colhead{RV$_{\rm TERRA}$} &
  \colhead{$\sigma_{\rm TERRA}$} &
  \colhead{RV$_{\rm CCF}$} &
  \colhead{$\sigma_{\rm CCF}$} &
  \colhead{S-index} &
  \colhead{$\sigma_{S}$}
  \\
  \colhead{[days]} &
  \colhead{[\ms]} &
  \colhead{[\ms]} &
  \colhead{[\ms]} &
  \colhead{[\ms]} &
  \colhead{\, } &
  \colhead{\, }
  }
\startdata
 2452990.814510 &   -0.88325 &    0.36747 &    0.84604 &    0.41236 &    0.30296 &    0.00201 \\
 2453002.833631 &   -1.67383 &    0.46082 &    0.26290 &    0.36177 &    0.29365 &    0.00160 \\
 2453051.749270 &    1.74874 &    0.70950 &    2.85012 &    0.42727 &    0.31230 &    0.00201 \\
 2453063.674457 &    2.03416 &    0.56129 &    2.97591 &    0.33228 &    0.29376 &    0.00156 \\
 2453759.772682 &   -1.61672 &    0.37845 &   -1.79124 &    0.20737 &    0.36436 &    0.00110 \\
 2454141.753872 &   -0.12664 &    0.37722 &   -0.04298 &    0.19197 &    0.44411 &    0.00111 \\
 2454422.852079 &    1.03102 &    0.31708 &    1.00156 &    0.22558 &    0.42707 &    0.00125 \\
 2454426.830209 &   -0.45662 &    0.21420 &   -0.06601 &    0.17664 &    0.42802 &    0.00097 \\
\enddata

\label{tab:rv_hd85512}
\end{deluxetable}

\begin{deluxetable}{lcccccc}  
\tablecolumns{5}

\tablecaption{RV measurements on
Barnard's star (GJ 699). Last two columns are the 
RV measurements obtained using the \textit{optimal} 
redmost part of the spectrum only 
($6000 < \lambda< 6800$ \AA, see Section 
\ref{sec:apertures}).
}

\tablehead{
  \colhead{JD } &
  \colhead{RV$_{\rm TERRA}$} &
  \colhead{$\sigma_{\rm TERRA}$} &
  \colhead{RV$_{\rm CCF}$} &
  \colhead{$\sigma_{\rm CCF}$} &
  \colhead{RV$_{\rm TERRA-Red}$} &
  \colhead{$\sigma_{\rm TERRA-Red}$} 
  \\
  \colhead{[days]} &
  \colhead{[\ms]} &
  \colhead{[\ms]} &
  \colhead{[\ms]} &
  \colhead{[\ms]} &
  \colhead{\, } &
  \colhead{\, }
  }
\startdata
 2454194.893890 &    0.00000 &    0.34651 &    0.57194 &    0.32376 &    0.00000 &    0.34651 \\
 2454196.883399 &    0.81911 &    0.35716 &    0.84845 &    0.30566 &    0.81911 &    0.35716 \\
 2454197.894815 &   -0.78946 &    0.33255 &   -0.38601 &    0.31239 &   -0.78946 &    0.33255 \\
 2454198.890257 &    1.50797 &    0.45543 &    2.70673 &    0.43272 &    1.50797 &    0.45543 \\
 2454199.918201 &    0.44577 &    0.32082 &    0.35208 &    0.26366 &    0.44577 &    0.32082 \\
 2454200.920621 &   -0.50497 &    0.31501 &   -0.27627 &    0.30062 &   -0.50497 &    0.31501 \\
 2454202.918690 &    0.85727 &    0.29046 &    1.12713 &    0.30629 &    0.85727 &    0.29046 \\
 2454227.843799 &    1.28708 &    0.35269 &    0.93223 &    0.34197 &    1.28708 &    0.35269 \\
\enddata

\label{tab:rv_barnard}
\end{deluxetable}

\begin{deluxetable}{lcccccc}  
\tablecolumns{5}

\tablecaption{RV measurements on
Kapteyn's star (GJ 191). Last two columns are the 
RV measurements obtained using the \textit{optimal} 
redmost part of the spectrum only 
($4670 < \lambda< 6800$ \AA, see Section 
\ref{sec:apertures}).
}

\tablehead{
  \colhead{JD } &
  \colhead{RV$_{\rm TERRA}$} &
  \colhead{$\sigma_{\rm TERRA}$} &
  \colhead{RV$_{\rm CCF}$} &
  \colhead{$\sigma_{\rm CCF}$} &
  \colhead{RV$_{\rm TERRA-Red}$} &
  \colhead{$\sigma_{\rm TERRA-Red}$} 
  \\
  \colhead{[days]} &
  \colhead{[\ms]} &
  \colhead{[\ms]} &
  \colhead{[\ms]} &
  \colhead{[\ms]} &
  \colhead{[\ms]} &
  \colhead{[\ms]} 
  }
\startdata
 2452985.741114 &    4.17510 &    1.03818 &    4.55727 &    0.86033 &    3.81838 &    0.83699 \\
 2452996.763214 &    2.60098 &    0.26086 &    2.77118 &    0.56030 &    2.70283 &    0.21930 \\
 2453337.800966 &   -1.04323 &    1.09789 &    0.12095 &    0.76332 &   -0.74766 &    0.90860 \\
 2453668.835375 &   -2.28865 &    0.59397 &   -3.79258 &    0.49284 &   -2.21699 &    0.54708 \\
 2453670.739478 &   -3.94631 &    0.58109 &   -4.59149 &    0.41933 &   -4.25875 &    0.56210 \\
 2454141.574447 &    0.03985 &    0.52542 &    0.72655 &    0.37960 &   -0.13958 &    0.50784 \\
 2454167.530533 &   -1.47217 &    0.50752 &   -2.15028 &    0.38433 &   -1.66431 &    0.47212 \\
 2454169.505216 &   -2.43451 &    0.47790 &   -2.29499 &    0.34512 &   -2.51582 &    0.48129 \\
\enddata

\label{tab:rv_gj191}
\end{deluxetable}

\begin{deluxetable}{lcccccc}  
\tablecolumns{5}

\tablecaption{RV measurements on
Proxima Cen (GJ 551). Last two columns are the 
RV measurements obtained using the \textit{optimal} 
redmost part of the spectrum only 
($4400 < \lambda< 6800$ \AA, see Section 
\ref{sec:apertures}).
}

\tablehead{
  \colhead{JD } &
  \colhead{RV$_{\rm TERRA}$} &
  \colhead{$\sigma_{\rm TERRA}$} &
  \colhead{RV$_{\rm CCF}$} &
  \colhead{$\sigma_{\rm CCF}$} &
  \colhead{RV$_{\rm TERRA-Red}$} &
  \colhead{$\sigma_{\rm TERRA-Red}$} 
  \\
  \colhead{[days]} &
  \colhead{[\ms]} &
  \colhead{[\ms]} &
  \colhead{[\ms]} &
  \colhead{[\ms]} &
  \colhead{[\ms]} &
  \colhead{[\ms]} 
  }
\startdata
 2453152.599771 &    0.37348 &    0.47936 &    0.79047 &    1.02223 &    0.37873 &    0.45254 \\
 2453202.584812 &   -2.97293 &    2.29999 &   -0.93952 &    2.25726 &   -2.86609 &    2.15906 \\
 2453203.537223 &    1.17250 &    0.68714 &    1.90963 &    1.04968 &    0.93396 &    0.61743 \\
 2453205.531679 &    1.16372 &    3.95272 &    7.52242 &    1.30372 &    1.87165 &    1.98058 \\
 2453207.505219 &    0.66168 &    1.92677 &   -0.91287 &    4.20013 &    0.29104 &    1.79977 \\
 2453207.510890 &    2.95883 &    2.08065 &    0.91052 &    3.97917 &    3.82038 &    1.82329 \\
 2453577.495602 &    3.73820 &    0.49688 &    1.54178 &    1.09223 &    3.60389 &    0.46023 \\
 2453809.893589 &   -2.17564 &    0.51088 &   -0.64866 &    0.74769 &   -2.11864 &    0.48677 \\
\enddata
\label{tab:rv_proxima}
\end{deluxetable}

\begin{deluxetable}{lcccccc}  
\tablecolumns{5}

\tablecaption{Differential RV and S-index measurements
on 34 epochs of GJ 676A}

\tablehead{
  \colhead{JD } &
  \colhead{RV$_{\rm TERRA}$} &
  \colhead{$\sigma_{\rm TERRA}$} &
  \colhead{RV$_{\rm CCF}$} &
  \colhead{$\sigma_{\rm CCF}$} &
  \colhead{S-index} &
  \colhead{$\sigma_{S}$}
  \\
  \colhead{[days]} &
  \colhead{[\ms]} &
  \colhead{[\ms]} &
  \colhead{[\ms]} &
  \colhead{[\ms]} &
  \colhead{\, } &
  \colhead{\, }
  }
\startdata
 2454167.897856 &  143.65593 &    1.45050 &  144.59273 &    0.96910 &    1.27632 &    0.01023 \\
 2454169.895854 &  140.76651 &    1.38811 &  141.45382 &    0.83020 &    1.29254 &    0.00959 \\
 2454171.904445 &  145.84144 &    1.52625 &  143.32531 &    0.94088 &    1.23147 &    0.01024 \\
 2454232.818013 &  142.10548 &    1.19654 &  143.81091 &    0.78189 &    1.34713 &    0.01004 \\
 2454391.491808 &  -23.38607 &    1.40115 &  -18.92497 &    0.76311 &    1.39087 &    0.00982 \\
 2454393.489934 &  -27.22405 &    1.31049 &  -24.68468 &    0.87658 &    1.42889 &    0.01102 \\
 2454529.900847 &  -99.67167 &    1.40944 &  -99.46562 &    1.03749 &    1.36065 &    0.01155 \\
 2454547.915016 &  -98.53057 &    1.46694 &  -96.80962 &    0.79384 &    1.40050 &    0.00987 \\
\enddata

\label{tab:GJ676A}
\end{deluxetable}

\begin{deluxetable}{lcccccc}  
\tablecolumns{5}

\tablecaption{Differential RV and S-index measurements
on $\epsilon$ Eridani.}

\tablehead{
  \colhead{JD } &
  \colhead{RV$_{\rm TERRA}$} &
  \colhead{$\sigma_{\rm TERRA}$} &
  \colhead{RV$_{\rm CCF}$} &
  \colhead{$\sigma_{\rm CCF}$} &
  \colhead{S-index} &
  \colhead{$\sigma_{S}$}
  \\
  \colhead{[days]} &
  \colhead{[\ms]} &
  \colhead{[\ms]} &
  \colhead{[\ms]} &
  \colhead{[\ms]} &
  \colhead{\, } &
  \colhead{\, }
  }
\startdata
 2452949.789541 &  -13.27658 &    0.20836 &  -26.59166 &    0.24191 &    0.41652 &    0.00078 \\
 2452949.790432 &  -12.55701 &    0.13163 &  -26.30259 &    0.23060 &    0.41713 &    0.00075 \\
 2452949.791323 &  -13.71360 &    0.19637 &  -27.35248 &    0.22330 &    0.41628 &    0.00077 \\
 2453001.598269 &   -2.21518 &    0.63673 &   -1.45770 &    0.26916 &    0.41976 &    0.00090 \\
 2453039.595541 &   -1.14426 &    0.57985 &  -13.96471 &    0.25574 &    0.42908 &    0.00093 \\
 2453039.597010 &   -1.65446 &    0.53546 &  -14.61557 &    0.23938 &    0.43211 &    0.00086 \\
 2453039.598365 &   -1.14070 &    0.61844 &  -14.48246 &    0.23418 &    0.43038 &    0.00086 \\
 2453279.824466 &    3.86346 &    0.33977 &    4.53516 &    0.20579 &    0.40336 &    0.00068 \\
\enddata

\label{tab:rv_epseri}
\end{deluxetable}

\begin{deluxetable}{lcccccc}  
\tablecolumns{5}

\tablecaption{Measurements on
HD 69830. Instead of the CCF RVs, RV using the redder part
of the spectrum (see Section \ref{sec:HD69830_wave}) are provided 
in this case.
}

\tablehead{
  \colhead{JD } &
  \colhead{RV$_{\rm TERRA}$} &
  \colhead{$\sigma_{\rm TERRA}$} &
  \colhead{RV$_{\rm TERRA-Red}$} &
  \colhead{$\sigma_{\rm TERRA-Red}$} &
  \colhead{S-index} &
  \colhead{$\sigma_{S}$}
  \\
  \colhead{[days]} &
  \colhead{[\ms]} &
  \colhead{[\ms]} &
  \colhead{[\ms]} &
  \colhead{[\ms]} &
  \colhead{\, } &
  \colhead{\, }
  }
\startdata
 2452939.874025 &    0.78514 &    0.44513 &   -0.44804 &    0.49607 &    0.14331 &    0.00049 \\
 2452943.876091 &    4.52766 &    0.54224 &    3.79835 &    0.50728 &    0.14377 &    0.00043 \\
 2452946.863070 &    1.66663 &    0.54780 &   -0.08756 &    0.62966 &    0.14381 &    0.00042 \\
 2452949.840269 &    5.06226 &    0.97553 &    1.66953 &    0.94835 &    0.13985 &    0.00070 \\
 2452949.842734 &    4.21729 &    0.92317 &    0.86545 &    0.95698 &    0.14056 &    0.00064 \\
 2452949.845246 &    3.97665 &    0.83998 &    2.04983 &    0.81015 &    0.13968 &    0.00065 \\
 2452949.849471 &    3.85089 &   12.32709 &    2.06592 &   42.48213 &    0.14061 &    0.00067 \\
 2452949.851937 &    5.97607 &   11.50930 &    1.61959 &   41.41179 &    0.14058 &    0.00066 \\
\enddata

\label{tab:rv_HD69830}
\end{deluxetable}

\end{document}